\documentclass[aps,showpacs,nofootinbib,floatfix]{revtex4}

\usepackage{graphicx}

\usepackage{amsmath, amssymb, bm, graphicx, graphics, color,mathrsfs,multirow,ulem}

\newcommand{\be}{\begin{equation}}
\newcommand{\ee}{\end{equation}}
\newcommand{\ben}{\begin{eqnarray}}
\newcommand{\een}{\end{eqnarray}}

\newcommand{\la}{{\lambda}}

\newcommand{\cL}{{\cal L}}

\newcommand{\na}{\nabla}

\newcommand{\tpe}{{\tilde p}}

\newcommand{\hI}{\hat I}
\newcommand{\hg}{\hat g}
\newcommand{\hR}{\hat R}
\newcommand{\hna}{\hat \nabla}

\newcommand{\tT}{\tilde T}

\newcommand{\zpsi}{\psi^{\ast}}

\begin{document}

\title{Dynamical Collapse of Charged Scalar Field in Phantom Gravity}
%%%%%%%%%%%%%%%%%%%%%%%%%%%%%%%%%%%%%%%%%%%%%%%%%%%%%%%%%%%%%%%%%%%%%%%%%%%%%%%%%%%%%%%%%%%%%%%%%%%%%%%%%%%%%%%
%\author{David Langlois}
%\affiliation{Institute of  \protect \\
%University \protect \\
%20-031 , France \protect \\
%david@tytan.umcs.lublin.fr \protect \\

%\author{Anna Borkowska}
%\email[]{aborkow@kft.umcs.lublin.pl}
%\author{Marek Rogatko}
%\email[]{rogat@tytan.umcs.lublin.pl}
%\email[]{rogat@kft.umcs.lublin.pl}
%\thanks{}
%\affiliation{Institute of Physics \protect \\
%Maria Curie-Sklodowska University \protect \\
%20-031 Lublin, pl.~Marii Curie-Sklodowskiej 1, Poland}

\author{Anna Nakonieczna and Marek Rogatko}
\affiliation{Institute of Physics \protect \\
Maria Curie-Sklodowska University \protect \\
20-031 Lublin, pl.~Marii Curie-Sklodowskiej 1, Poland \protect \\
aborkow@kft.umcs.lublin.pl \protect \\
rogat@kft.umcs.lublin.pl \protect \\
marek.rogatko@poczta.umcs.lublin.pl}

\author{ Rafa{\l} Moderski}
\affiliation{Nicolaus Copernicus Astronomical Center \protect \\
Polish Academy of Sciences \protect \\
00-716 Warsaw, Bartycka 18, Poland \protect \\
moderski@camk.edu.pl }

%%%%%%%%%%%%%%%%%%%%%%%%%%%%%%%%%%%%%%%%%%%%%%%%%%%%%%%%%%%%%%%%%%%%%%%%%%%%%%%%%%%%%%%%%%%%%%%%%%%%%%%%
%%%%%%%%%%%%%%%%%%%%%%%%%%%%%%%%%%%%%%%%%%%%%%%%%%%%%%%%%%%%%%%%%%%%
\date{\today}
%\pacs{04.30.Nk, 04.40.-b}

%%%%%%%%%%%%%%%%%%%%%%%%%%%%%%%%%%%%%%%%%%%%%%%%%%%%%%%%%%%%%%%%%%%%%%
\begin{abstract}
We investigated the problem of the dynamical collapse of a self-gravitating complex charged scalar field
in Einstein-Maxwell-dilaton theory with a phantom coupling for the adequate fields in the system under consideration.
We also considered two simplifications of it, i.e., the separate collapses of phantom Maxwell and phantom scalar fields 
under the influence of Einstein gravity.
One starts with the regular spacetime and leads the evolution through the formation of the horizons and the 
final singularity. We discuss the structures of spacetimes emerging in the process of the dynamical collapse 
and comment on the role of the considered fields in its course.

\end{abstract}
%%%%%%%%%%%%%%%%%%%%%%%%%%%%%%%%%%%%%%%%%%%%%%%%%%%%%%%%%%%%%%%%%%%%%%%

\maketitle
%%%%%%%%%%%%%%%%%%%%%%%%%%%%%%%%%%%%%%%%%%%%%%%%%%%%%%%%%%%%%%%%%%%%%%%%%

\section{Introduction}
Phantom gravitating fields emerge with a kinetic term with the {\it wrong sign}. It turns out that this fact
implies that they are coupled repulsively to gravity. These fields could form {\it ghost condensation}
at the quantum level, which in turn might lead to the modification of gravity at the infrared limit
\cite{ark04}. On the other hand, the present observations of the Universe reveal the fact that
it may be dominated by an {\it exotic} kind of fields with negative pressure causing the acceleration.
Type $Ia$ supernovae (Sn Ia) provide a precise calibrated {\it standard candle} with which one can probe
the expansion of the Universe on large scales. In the late 1990s the independent surveys for distant
Sn Ia announced that the high-redshifted supernovae of this type appeared about $40$ percent fainter or
equivalently more distant than expected in a flat, matter-dominated Universe \cite{accuniv}.
These researches were combined with other concerning the composition of the Universe such as 
cosmic microwave background (CMB) observations \cite{ben03} or analyzes of barionic acoustic oscillations (BAO) \cite{eis05}. 
Altogether they do not exclude the possibility that the {\it exotic} matter mentioned above consists of 
the fields with the so-called super-negative pressure, i.e., when $p<-\rho$ ($p$ and $\rho$ stand for 
the pressure and density of matter, respectively) \cite{cal02}. Such fields are dubbed phantom.
\par
The presence of this kind of fields which contribute negatively to the total energy, i.e.,
when the violation of the null energy condition takes place, gives rise to the interesting possibilities
of new solutions of Einstein equations coupled to phantom field. Namely, in Ref.\cite{ell73} the wormhole
solutions with phantom scalar field were studied both analytically and numerically. Phantom field has also 
a direct consequence for black hole physics.
In Ref.\cite{gib96} in theories where gravity coupled to Maxwell and dilaton fields in such
a way that the kinetic energy for either or both of the fields in question was allowed to be negative,
black hole solutions were obtained. Among all it was found that regular black hole solutions with
zero or negative mass could be provided. On the other hand, 
cold black holes with a multiply degenerate 
event horizon and an infinite horizon area were considered in \cite{cle99}. 
The new class of static, spherically symmetric solutions in Einstein-Maxwell-dilaton theory with phantom
coupling for the dilaton and/or Maxwell field leading to black holes with single or multiple horizons were
elaborated in Ref.\cite{cle09}. The static multicenter solutions of 
phantom Einstein-Maxwell-dilaton theory and gravitating $\sigma$-models obtained via dimensional reduction
of phantom Einstein-Maxwell, phantom Kaluza-Klein and phantom Einstein-Maxwell-dilaton-axion theories were 
discussed in \cite{cle11}.
%%%%%%%%%%%%%%%%%%%%%%%%%%%%%%%%%%%%%%%%%%%%%%%%%%%%%%%%%%%%%%%%%%%%%%%%%%%%%%%%%%%%%%%%%%%%%%%%%%%%%%
\par
On the other hand, the gravitational collapse 
and the {\it no-hair} conjecture or its mathematical formulation black hole uniqueness theorem,
stating that stationary axisymmetric solution of Einstein-Maxwell equations 
relaxes to Kerr-Newman spacetime and is characterized 
by black hole mass, charge and angular momentum \cite{uniq}, attracted attention for many decades of researches.
A lot of progress has been also made during last years to answer the basic question
underlying the theoretical studies of black hole interiors. 
Quite different description of the inner black hole singularity emerged \cite{his81}-\cite{is}.
A full nonlinear description of the inner structure
of black holes was presented in Ref.\cite{gne93}. Then, in \cite{ham96} the spherically symmetric 
collapse of massless scalar field was numerically examined. It was shown that the field
dispersed to infinity or collapsed to a black hole. The aforementioned researches were also broaden
to the case of the nonlinear evolution of the neutral scalar field on a background spacetime of charged black hole
\cite{bra95}. Ref.\cite{aya97} was devoted to the numerical solution of spherically symmetric
semi-classical scalar field collapse. The addition of the effective energy momentum tensor
describing evaporation of black hole enabled to consider the influence of this process on a critical
phenomena present in classical collapse.
In \cite{hod98} the occurrence of the {\it mass inflation}
phenomenon during a dynamical charged gravitational collapse was shown. As far as {\it mass inflation}
is concerned, it was shown in \cite{is} that in a realistic RN black hole the Cauchy horizon was unstable
due to the fact that it developed a weak null singularity caused by the local mass parameter growing indefinitely
large.
\par
The case of the dynamical collapse of spherically symmetric shell of the charged massless scalar field
was elaborated in Ref.\cite{ore03}. An external RN spacetime and the inner spacetime bounded by singularity
on the Cauchy horizon were obtained from numerical calculations. 
In order to figure out pair creation process in strong electric field in the dynamical collapse
of electrically charged massless scalar field, the dynamical formation of evaporating spherically symmetric
charged black hole emitting Hawking radiation and the behaviour of complex scalar field, gauge field
and the normalized energy momentum tensor during the process in question, numerical approaches
were performed \cite{sor01}-\cite{hwa11}. These researches were supplemented by the studies 
of the behaviour of Brans-Dicke field \cite{hwa10} during gravitational collapse of matter. 
\par
Apart from studying the full gravitational collapse on flat background spacetime, the wide range of simulations 
of an accretion onto an existing black hole was also carried out. The most interesting for us 
{\it exotic} matter accreting onto the Schwarzschild \cite{gon09} or Reissner-Nordstr\"{o}m black hole \cite{dor10}
was modeled as a free scalar field with the opposite sign in its energy momentum tensor.
\par
Because of the fact that 
the uniqueness theorem for the low-energy string black holes is 
quite well established \cite{len},
the implication of the superstring theory on the dynamical gravitational collapse 
of charged scalar field was studied in \cite{bor11}. Numerical studies revealed that the dynamical collapse
in question led to the formation of Schwarzschild-like spacetimes in case of non-zero coupling between 
the charged scalar field and the fields emerging from the superstring theory.
\par
As was mentioned above, the presence of phantom fields as well as phantom coupling of the ordinary
fields, gives rise to new possibilities for solutions of {\it generalized} Einstein equations.
Motivated by the aforementioned analytical and numerical researches we shall consider the problem
of the full nonlinear dynamical collapse of a complex scalar field when gravitational interactions take forms of
Einstein-{\it phantom Maxwell}~(E$\overline{M}$),
Einstein-{\it phantom Maxwell}-dilaton~(E$\overline{M}D$), Einstein-Maxwell-{\it phantom dilaton}~($EM\overline{D}$),
and Einstein-{\it phantom Maxwell}-{\it phantom dilaton}~(E$\overline{MD}$) theories. 
We will also consider the evolution in Einstein-{\it phantom dilaton}~(E$\overline{D}$) theory, 
namely the collapse of phantom scalar field in the regime of Einstein gravity.
\par
Our paper is organized as follows. In Sec.II
we derive equations of motion describing the gravitational collapse of a shell of charged scalar field in
Einstein-Maxwell-dilaton theory with {\it phantom} coupling. Sec.III is connected with the numerical scheme applied
in our investigations. In Sec.IV we discuss obtained results, while in Sec.V we present the conclusions of 
our investigations.

%%%%%%%%%%%%%%%%%%%%%%%%%%%%%%%%%%%%%%%%%%%%%%%%%%%%%%%%%%%%%%%%%%%%%
\section{Phantom Einstein-Maxwell-dilaton theory}
The main aim of our researches will be to study the dynamical collapse of a complex scalar field in the string-inspired background
with {\it phantom} coupling of the fields in the theory.
Because of the fact that the coupling constant of dilaton field to the complex scalar field is unknown, the 
considered action is provided in the form written in the {\it string frame} as follows:
\be
\hI = \int d^{4} x \sqrt{- \hg} \left [ e^{- 2 \phi}
\left (
\hR - 2 \xi_1 \left ( \hna \phi \right )^2 + e^{ 2 \alpha \phi} \cL \right )
\right ],
\label{a1}
\ee
where $\phi$ denotes the dilaton field, $\alpha$ stands for the coupling constant between dilaton 
and charged scalar field $\psi$ described by the Lagrangian $\cL$ of the form as
\be
\cL = - {1 \over 2} \left (
\hna_{\alpha} \psi + i e A_{\alpha} \psi \right ) \hg^{\alpha \beta}
\left (
\hna_{\beta} \zpsi - i e A_{\beta} \zpsi \right ) - \xi_2 F_{\mu \nu}
F^{\mu \nu} .
\ee
$\xi_1$ and $\xi_2$ are dilaton-gravity and $U(1)$-gauge field-gravity coupling constants. 
In the normal case of Einstein-Maxwell-dilaton-complex scalar field system they are equal to $1$. On the contrary,
in {\it phantom} case we have $\xi_1 = -1$ and/or $\xi_2 = -1$. After performing the conformal transformation,
the action may be rewritten in the so-called {\it Einstein frame}.
It yields 
\be
I = \int d^{4} x \sqrt{-g} \bigg[
R - 2 \xi_1 ( \na \phi )^{2} + e^{2 \alpha \phi + 4 \phi} \cL
(\psi, \zpsi, A, e^{2 \phi} g_{\alpha \beta}) \bigg].
\ee
Taking variations with respect to the adequate fields in the theory in question, we obtain the following
equations of motion:
\ben
\na^{2} \phi &-&
{\alpha + 1 \over 4 \xi_1}~e^{2 \phi (\alpha + 1)}  \bigg( 
\na_{\beta} \psi + i e A_{\beta} \psi \bigg) 
\bigg( \na^{\beta} \zpsi - i e A^{\beta} \zpsi \bigg)
- {1 \over 2} {\xi_2 \over \xi_1} \alpha
e^{ 2 \alpha \phi} F^{2} = 0,  \label{aaa} \\
 \na_{\mu} \bigg( e^{ 2 \alpha  \phi} F^{\mu \nu} \bigg) 
&+& {e^{2 \phi (\alpha + 1)}\over 4 \xi_2} \bigg[
i e \zpsi \bigg( \na^{\nu} \psi + i e A^{\nu} \psi \bigg)
- i e \psi \bigg( \na^{\nu} \zpsi - i e A^{\nu} \zpsi \bigg)
\bigg] = 0, \label{bbb}\\
\na^{2} \psi &+& i e A^{\beta} \bigg(
2 \na_{\beta} \psi + i e A_{\beta} \psi \bigg)
+ i e \na_{\delta} A^{\delta} \psi = 0, \label{ccc} \\
\na^{2} \zpsi &-& i e A^{\beta} \bigg(
2 \na_{\beta} \zpsi - i e A_{\beta} \zpsi \bigg)
- i e \na_{\delta} A^{\delta} \zpsi = 0, \label{c1c1c1} \\
G_{\mu \nu} &=& T_{\mu \nu}(\phi, F, \psi, \zpsi, A).
\een
The energy momentum tensor $T_{\mu \nu}(\phi, F, \psi, \zpsi, A)$ is given by
\be
T_{\mu \nu}(\phi, F, \psi, \zpsi, A) = e^{2 \phi( \alpha + 1)}
\tT_{\mu \nu}(\psi, \zpsi, A) + T_{\mu \nu}(F, \phi),
\label{ten}
\ee
where $\tT_{\mu \nu}(\psi, \zpsi, A)$ implies
\ben
\tT_{\mu \nu}(\psi, \zpsi, A) &=& 
{1 \over 4} \bigg[
ie ~\psi \bigg( 
A_{\mu}~\na_{\nu} \zpsi + A_{\nu}~\na_{\mu} \zpsi \bigg) - 
ie~\zpsi \bigg( 
A_{\mu}~\na_{\nu}\psi + A_{\nu}~\na_{\mu}\psi \bigg) \bigg]
\\ \nonumber
&+&
{1 \over 4} \bigg(
\na_{\mu} \psi~ \na_{\nu} \zpsi + \na_{\mu} \zpsi~ \na_{\nu} \psi \bigg) +
{1 \over 2}~e^2 A_{\mu}~ A_{\nu} \psi~ \zpsi +
{1 \over 2}~\tilde {\cal L}(\psi, \zpsi, A) g_{\mu \nu}.
\een
On the other hand, 
the explicit form of the Lagrangian $\tilde {\cal L}(\psi, \zpsi, A)$ is provided by the expression
\be
\tilde {\cal L}(\psi, \zpsi, A) = - {1 \over 2} \bigg(
\na_{\beta} \psi + i e A_{\beta} \psi \bigg)~
\bigg( \na^{\beta} \zpsi - i e A^{\beta} \zpsi \bigg).
\ee
The energy momentum tensor for $U(1)$-gauge and dilaton fields yields
\be
T_{\mu \nu}(F, \phi) = \xi_2 e^{2 \alpha \phi} \bigg(
2 F_{\mu \rho} F_{\nu}{}{}^{\rho} - {1 \over 2} g_{\mu \nu} F^2 \bigg )
+ \xi_1 \bigg( 2 \na_{\mu} \phi \na_{\nu} \phi - g_{\mu \nu} (\na \phi)^2 \bigg).
\label{ten-2}
\ee

%%%%%%%%%%%%%%%%%%%%%%%%%%%%%%%%%%%%%%
In our research we shall start with the regular initial spacetime at approximately past null infinity, 
compute the formation of the horizons and follow the evolution of the dynamically formed black hole to the central singularity.
One chooses $(2+2)$-spherically symmetric double null coordinate system. It turned out that any coordinate
gauge transformation of the form $u \rightarrow f(u)$ and $v \rightarrow g(v)$ will preserve the null character of the
retarded and advanced coordinates. The line element written in the aforementioned coordinates is provided by
\cite{chr93}:
\be
ds^2 = - a(u, v)^2 du dv + r^2(u, v) d \Omega^2.
\label{m}
\ee
The assumptions about spherical symmetry of the problem imply that the only non-vanishing components of
the $U(1)$-gauge strength field are $F_{uv}$ and $F_{vu}$.
Moreover, another restriction on the gauge potential can be implemented. We can consider $A_{u}$ or $A_{v}$.
It turned out \cite{bor11} that one can restrict his attention to the only one gauge potential due to the fact
of the gauge freedom $A_{u} \rightarrow A_{u} + \na_{u} \theta$, where $\theta = \int A_{v}dv$. Consequently,
we are left with the only one component of the gauge field being the function of the retarded and advanced coordinates.\\
Defining the quantity like in Ref.\cite{ore03}:
\be
Q = 2 {A_{u, v}~r^2 \over a^2},
\label{charge}
\ee 
enables us to separate $v$-component of the 
second order partial differential {\it generalized} Einstein-Maxwell equations (\ref{bbb}) given by
\be
\bigg[ {2 e^{2 \alpha \phi}~r^2~A_{u,v} \over a^2} \bigg]_{,v} +
{r^2~e^{2 \phi (\alpha + 1)} \over 4 \xi_2 }~ie~  \bigg(
\zpsi~\psi_{,v} - \psi~\zpsi_{,v} \bigg) = 0
\ee
into much simpler first order differential equations for $A_{u}$
\be
A_{u, v} - {Q a^{2} \over 2 r^{2}} = 0
\label{p}
\ee
and for $Q$, which in turn yields
\be
Q_{,v} + 2~ \alpha~ \phi_{,v}~ Q + {i e r^2 \over 4 \xi_2} e^{2 \phi}
\bigg( \zpsi \psi_{,v} - \psi \zpsi_{,v} \bigg) = 0. 
\label{l}
\ee
%%%%%%%%%%%%%%%%%%%%%%%%%%%%%%%%%%%%%%%%%%%%%%%%%%%%%%%%%%%%%%%%%%%%%%%%%%%%%%%%%%%%%%%%%%%%%%%%%%%%%%
On the other hand, Eq.(\ref{aaa}) has the explicit form as follows:
\be
r_{,u} \phi_{,v} + r_{,v} \phi_{,u} + r \phi_{,uv} -
{(\alpha + 1) \over 8 \xi_1}~e^{2 \phi (\alpha + 1)}~ r~ \bigg[
\psi_{,u} \zpsi_{,v} + \psi_{,v} \zpsi_{,u} + i e~ A_{u}
\bigg( \psi \zpsi_{,v} - \zpsi \psi_{,v} \bigg ) \bigg ]
- \alpha {\xi_2 \over \xi_1}
{a^2~Q^2~ e^{ 2 \alpha \phi}
\over 4 r^3} = 0.
\ee
An inspection of the relations for the complex scalar field (\ref{ccc})-(\ref{c1c1c1})
reveals that one gets the following:
\ben
r_{,u} \psi_{,v} + r_{,v} \psi_{,u} + r \psi_{,uv} + i e~ r~ A_{u}~ \psi_{,v}
+ i e~ r_{,v}~ A_{u}~ \psi + {i e~ Q~ a^2 \over 4 r} \psi = 0, \\
r_{,u} \zpsi_{,v} + r_{,v} \zpsi_{,u} + r \zpsi_{,uv} - i e~ r~ A_{u}~ \zpsi_{,v}
- i e~ r_{,v}~ A_{u}~ \zpsi - {i e~ Q~ a^2 \over 4 r} \zpsi = 0.
\een
Furthermore, the adequate components of the Einstein tensor and the stress-energy tensor for the underlying
theory reveal the set of equations
\ben
{2 a_{,u}~ r_{,u} \over a} - r_{,u u}
&=& \xi_1 r~ \phi_{,u}^2 +
{r~e^{2 \phi (\alpha + 1)} \over 4} \bigg[
\psi_{,u} \zpsi_{,u} + i e~ A_{u}~ \bigg(
\psi~\zpsi_{,u} - \zpsi~\psi_{,u} \bigg ) + e^2 ~A_{u}^2 \psi~ \zpsi 
\bigg], \\
{2 a_{,v}~ r_{,v} \over a} - r_{,vv} &=& 
\xi_1 r~ {\phi_{v}}^2 + {1 \over 4}~r~ e^{2 \phi (\alpha + 1)}~ \psi_{,v}~ \zpsi_{,v}, \\
{a^2 \over 4r} + {r_{,u}~ r_{v} \over r} +  r_{,uv} &=& \xi_2 {e^{2 \alpha \phi}~a^2~Q^2 \over 4 r^3}, \\
{a_{,u} a_{,v} \over a^2}
- {a_{,uv} \over a} - {r_{,uv} \over r} &=&
\xi_2 {Q^2~ e^{2 \alpha \phi}~ a^2 \over 4 r^4} + \xi_1 \phi_{,u} \phi_{,v} 
+ {e^{2 \phi(\alpha + 1)} \over 8} \bigg [
\psi_{,u}~ \zpsi_{,v} + \zpsi_{,u}~ \psi_{,v}
+ i e~ A_{u}~ \bigg (\psi~ \zpsi_{v} - \zpsi~\psi_{,v} \bigg) 
\bigg].
\een
To proceed further, one introduces the new auxiliary variables written in the following forms:
\ben
c = \frac{a_{,u}}{a}, \qquad d = \frac{a_{,v}}{a},
\qquad f = r_{,u}, \qquad g = r_{,v},  \nonumber\\
s = \psi,  \qquad  p  =  \psi_{,u}, \qquad q  =  \psi_{,v}, \qquad \beta  =  A_u,\\
k = \phi, \qquad x = \phi_{,u}, \qquad y  = \phi_{,v}.  \nonumber
\label{eqn:substitution}
\een
Moreover, the additional quantities take the forms as
\ben
\la & \equiv & \frac{a^2}{4} + f g,
\label{eqn:lambda-definition}\\
\mu & \equiv & f q + g p,
\label{eqn:mi-definition}\\
\eta & \equiv & g x + f y.
\label{eqn:eta-definition}
\een
It can be easily seen that instead of 
taking into account two complex fields $\psi$ and $\zpsi$, one can introduce two real fields.
Namely, they satisfy the relations $\psi = \psi_{1} + i~\psi_{2}$ and $\zpsi = \psi_{1} - i~\psi_{2}$. 
Under these assumptions we arrive at                 
\ben \label{eqn:complex-functions-s-p-q}
s &=& s_1 + i~s_2, \qquad p  =  p_1 + i ~p_2, \qquad q  =  q_1 + i ~q_2, \\ \label{eqn:complex-function-mu}
\mu & = & \mu_1 + i ~\mu_2, \qquad
\mu_1 = f q_1 + g p_1, \qquad \mu_2  =  fq_2 + gp_2.
\label{eqn:complex-functions-mu1-mu2}
\een
Just, the system of the second order partial differential 
equations can be rewritten in the form of the first order one.
It can be checked that we finally arrive at the set of the first order differential equations written as follows:
\ben \label{eqn:P1-2}
& P1: & a_{,u} - a~c  =  0,\\
& P2: & a_{,v}- a~d =  0,\\
& P3: & r_{,u}- f  =  0,\\
& P4: & r_{,v} - g  =  0,\\
& P5_{_{\left(Re\right)}}: & s_{1,u} - p_1  =  0,\\
& P5_{_{\left(Im\right)}}: & s_{2,u} - p_2  =  0,\\
& P6_{_{\left(Re\right)}}: & s_{1,v} - q_1  =  0,\\
& P6_{_{\left(Im\right)}}: & s_{2,v} - q_2  =  0,\\
& P7: & k_{,u} - x  =  0,\\
& P8: & k_{,v} - y  =  0,\\
& E1: & f_{,u} - 2~c~f + \xi_1 r~x^2+\frac{1}{4}~r~e^{2k\left(\alpha+1\right)}
\Big[p_1^{\: 2} + p_2^{\: 2} + 2~e~\beta~\left(s_1~p_2 - s_2~p_1 \right)
+ e^2~\beta^2~\left(s_1^{\: 2} + s_2^{\: 2} \right) \Big]  =  0,\\
& E2: & g_{,v} - 2~d~g + \xi_1 r~y^2 + \frac{1}{4}~r~e^{2k\left(\alpha+1\right)}~\left(q_1^{\: 2} + q_2^{\: 2}\right)
= 0,\\
& E3^{\left(1\right)}: & f_{,v} + \frac{\la}{r} - \xi_2 e^{2\alpha k}~\frac{Q^2a^2}{4~r^3}  =  0,\\
& E3^{\left(2\right)}: & g_{,u} + \frac{\la}{r} - \xi_2 e^{2\alpha k}~\frac{Q^2a^2}{4~r^3}  =  0,\\
& E4^{\left(1\right)}: & c_{,v} - \frac{\la}{r^2} + \xi_1 x~y + \frac{1}{4}~e^{2k\left(\alpha+1\right)}
\Big[p_1~q_1 + p_2~q_2 + e~\beta~\left( s_1~q_2 - s_2~q_1 \right) \Big] + \xi_2 e^{2\alpha k} \frac{Q^2a^2}{2~r^4}  = 0,\\
& E4^{\left(2\right)}: & d_{,u} - \frac{\la}{r^2} + \xi_1 x~y
+ \frac{1}{4}~e^{2k\left(\alpha+1\right)}
\Big[p_1~q_1 + p_2~q_2 + e~\beta~\left( s_1~q_2 - s_2~q_1 \right) \Big] + \xi_2 e^{2\alpha k}\frac{Q^2a^2}{2~r^4}  =  0,\\
& S_{_{\left(Re\right)}}^{\left(1\right)}: & rp_{1,v} + \mu_1 - e~ r~\beta~ q_2 - e~ s_2~\beta~ g -
e~ s_2~\frac{Qa^2}{4r}  =  0,\\
& S_{_{\left(Im\right)}}^{\left(1\right)}: & rp_{2,v} + \mu_2 + e~ r~\beta~ q_1 + e~ s_1~\beta~ g + e~ s_1~\frac{Qa^2}{4r}
= 0,\\
& S_{_{\left(Re\right)}}^{\left(2\right)}: & rq_{1,u} + \mu_1 - e~ r~\beta~ q_2 - e~ s_2~\beta~ g - e~ s_2~\frac{Qa^2}{4r}
= 0,\\
& S_{_{\left(Im\right)}}^{\left(2\right)}: & rq_{2,u} + \mu_2 + e~ r~\beta~ q_1 + e~ s_1~\beta~ g + e~ s_1~\frac{Qa^2}{4r}
= 0,\\
& D^{\left(1\right)}: & rx_{,v} + \eta - \frac{\alpha+1}{4 \xi_1}~r~e^{2k\left(\alpha+1\right)}
\Big[p_1~q_1 + p_2~q_2 + e~\beta~\left(s_1~q_2 - s_2~q_1 \right) \Big] - \alpha {\xi_2 \over \xi_1} e^{2\alpha k}~\frac{Q^2a^2}{4~r^3}  =  0,\\
& D^{\left(2\right)}: & r~y_{,u} + \eta - \frac{\alpha+1}{4 \xi_1}~r~e^{2k\left(\alpha+1\right)}
\Big[p_1~q_1 + p_2~q_2 + e~\beta~\left( s_1~q_2 - s_2~q_1 \right) \Big] - \alpha {\xi_2 \over \xi_1} e^{2\alpha k}\frac{Q^2a^2}{4~r^3}  =  0,\\
& M1: & \beta_{,v} - \frac{Qa^2}{2~r^2}  =  0,\\
& M2: & Q_{,v} + 2~\alpha~ y~Q - {1 \over 2 \xi_2}~e^{2k}~e~ r^2 \left( s_1~q_2 - s_2 ~q_1 \right)  =  0.
\label{eqn:M2-2}
\een
%%%%%%%%%%%%%%%
\par
Let us introduce the quantity of the physical interest.
Namely, we define the mass function provided by the relation
\be
m(u,v) = {r \over 2}~\bigg( 1 + {4 ~r_{,u}~r_{,v} \over a^2} \bigg) =
{r \over 2}~\bigg( 1 + {4 \over a^2}~f~g \bigg) .
\label{mhaw}
\ee
It represents the Hawking mass, i.e., the mass included in a sphere of the radius $r(u,v)$.

The characteristics of all the evolutions under consideration are listed in Table \ref{tab:evolutions-characteristics}. 
As was explained in detail in \cite{cle11} the structures of solutions in $EMD$ as well as in {\it phantom}-$EMD$ theories may be regarded as equilibrium states resulting from the competition and/or cooperation among gravitational, 
electromagnetic and dilatonic forces. The nature of gravitational force is always attractive. The other two forces 
are either attractive or repulsive depending on whether they are {\it phantom} or not. 
It should be kept in mind that in our case we deal with one more constituent of the system under consideration, namely the complex scalar field. Its influence on spacetime structures is uniform in all considered cases, because we do not allow it to be {\it phantom}. We shall comment on its role during the collapse after a detailed analysis of spacetime structures.

\begin{table}
\caption{Characteristics of evolutions under consideration. The overlining indicates {\it phantom} field. Parameters $\tpe_k$ and $\tpe_s$ denote amplitudes of dilaton and electrically charged scalar fields, respectively.}
\begin{ruledtabular}
\begin{tabular}{c|c|c|c|c|c|c}
Type of evolution & $\xi_1$ & $\xi_2$ & $\tpe_k$ & $\tpe_s$ & $e$ & $\alpha$\\
\hline
& & & & & &\\
$E\overline{M}$ & $-$ & $-1$ & $-$ & $\neq 0$ & $\neq 0$ & $-$\\
& & & & & &\\
$E\overline{D}$ & $-1$ & $-$ & $\neq 0$ & $-$ & $-$ & $-$\\
& & & & & &\\
$E\overline{M}D$ & $+1$ & $-1$ & $\neq 0$ & $\neq 0$ & $\neq 0$ & $0,\: - 1$\\
& & & & & &\\
$EM\overline{D}$ & $-1$ & $+1$ & $\neq 0$ & $\neq 0$ & $\neq 0$ & $0,\: - 1$\\
& & & & & &\\
$E\overline{MD}$ & $-1$ & $-1$ & $\neq 0$ & $\neq 0$ & $\neq 0$ & $0,\: - 1$\\
& & & & & &\\
\end{tabular}
\end{ruledtabular}
\label{tab:evolutions-characteristics}
\end{table}

%%%%%%%%%%%%%%%%%%%%%%%%%%%%%%%%%%%%%%%%%%%%%%%%%%%%%%%%%%%%%%%%%%%%%%%%%%%%%%%%%%%%%%%%%%%%%%%%%%%%%%%%%%%%%%%%%%%%%%%%%%%%%%%

%%%%%%%%%%%%%%%%%%%%%%%%%%%%%%%%%%%%%%%%%%%%%%%%%%%%%%%%%%%%%%%%%%%%%%%%%%%%%%%%%%%%%%%%%%%%%%%%%%%%%%%%%%%%%%%%%%%%%%%%%%%%%%
%%%%%%%%%%%%%%%%%%%%%%%%%%%%%%%%%%%%%%%%%%%%%%%%%%%%%%%%%%%%%%%%%%%%%%%%%%%%%%%%%%%%%%%%%%%%%%%%%%%%%%%%%%%%%%%%%%%%%%%%%%%%%%%
\section{Numerical computations}
Unfortunately, a detailed analytical investigations of fully nonlinear gravitational dynamics described by
the set of the conjugate equations (\ref{eqn:P1-2})-(\ref{eqn:M2-2}) 
are impossible. Therefore, numerical methods ought to be used in 
order to draw conclusions about the structure of spacetime emerging in the process of
the dynamical collapse.

In the present paper we used a numerical algorithm containing an adaptive mesh 
refinement described in \cite{bor11}. It was adjusted for the current issue by taking the parameters 
$\xi_1$ and $\xi_2$ into account. The manner of setting the boundary and initial conditions 
for the considered equations remained unchanged in comparison to Ref.\cite{bor11}. The one-parameter 
families representing the initial profiles of the field functions used for simulations in the present paper 
are listed in Table \ref{tab:initial-profiles}. The $\left(f_D\right)$-family refers to the dilaton field 
with the family constants $c_1=1.3$ and $c_2=0.21$. The $\left(f_S\right)$-family describes the 
electrically charged scalar field. The family constant $v_f=7.5$ and the parameter determining the amount of 
the initial electric charge $\delta=\frac{\pi}{2}$ (maximal electric charge) \cite{hwa10}. In both cases 
the free family parameter is denoted by $\tpe$
with a subscript corresponding to the particular field. Such choice of 
the initial conditions is physically representative, because the considered collapse is universal, i.e., its course and results do not depend on the types of initial profiles.

There are no analytical solutions neither to the problem in question nor to any of simplified 
versions of it (apart from the extremely simplified, trivial case of empty spacetime). Thus, 
checking the correctness of numerical code has to be based on indirect methods.

The three tests used for proving the credibility of the code and justifications for carrying them out, 
are widely discussed in \cite{bor11}. As was reported in the 
aforementioned paper, the results obtained for the different integration steps on 
the non-adaptive grid display satisfying agreement of an order of $0.01\%$. Moreover, the code 
displays the linear convergence as is expected for the applied algorithm and errors decrease with 
increasing grid density. We reached the same conclusions for the modified code for all possible combinations 
of $\xi_{1,2}$ shown in Table \ref{tab:evolutions-characteristics}. Since the figures depicting 
them are analogous to the ones presented in the previous paper, we decided not to present them here.

The issue of mass (\ref{mhaw}) and charge (\ref{charge}) conservation was also analyzed. We obtained 
profiles of $m$ and $Q$ versus $u$ along ingoing null rays specific to all types of evolutions 
under consideration. The values of the advanced time were chosen in such a way that the null 
rays did not intersect any horizons for the small values of the retarded time. The obtained 
profiles appeared to be in a qualitative agreement with the ones presented in \cite{bor11}. For the evolutions 
leading to the formation of a black hole the mass was conserved within $7\%$ and the electric charge 
within $4.5\%$ in regions apart from the vicinities of the apparent horizons. The fluctuations were 
significant there due to the outgoing fluxes, which appear during the refraction of matter on the potential 
barriers near the gravitational radius of the collapsing shell \cite{ore03}. Mass and charge 
profiles for non-singular spacetimes also displayed expected behaviour. The values tended 
towards zero for the reason that evolving matter bounced off the 
non-singular $r=0$ and moved towards future null infinity leaving empty spacetime \cite{ham96}.

\begin{figure}[p]
\includegraphics[scale=0.5]{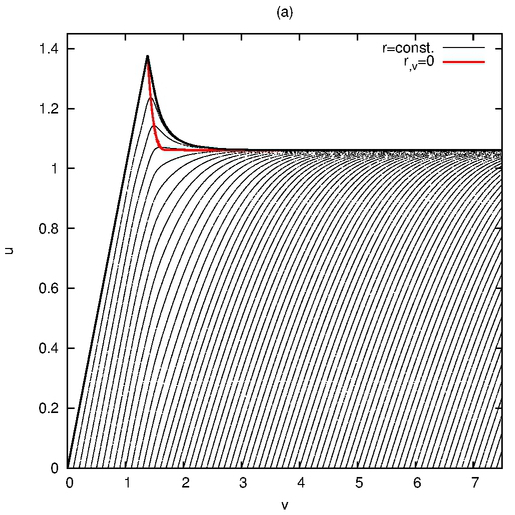}
\includegraphics[scale=0.5]{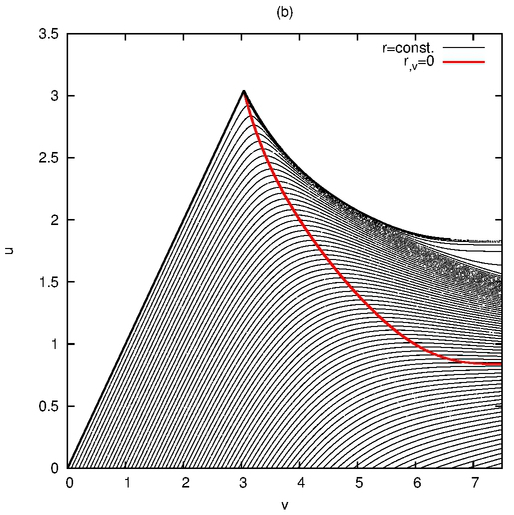}
\caption{(color online). Penrose diagrams of the dynamical (a) Schwarzschild and (b) 
Reissner-Nordstr\"{o}m spacetimes obtained for the field amplitudes equal to $\tpe_k=0.1$ and $\tpe_s=0.6$, respectively.}
\label{fig:RNS}
\end{figure}

Next, we examined the simplified versions of the problem in question, namely the non-{\it phantom} 
($\xi_1=\xi_2=1$) collapses of 
neutral and electrically charged scalar fields leading to Schwarzschild and 
Reissner-Nordstr\"{o}m spacetimes, respectively. 
The results are consistent with those published, e.g., in Refs.\cite{ham96,hod98,ore03}.\\
The dynamical Schwarzschild spacetime corresponding to the outcome of the collapse of a neutral scalar field \cite{ham96} in fact results from the evolution of dilaton field under the influence of gravitation ($ED$-collapse). 
Its structure for $\tpe_k=0.1$ is shown in Fig.\ref{fig:RNS}a. There is one apparent horizon visible in the spacetime, namely $r_{,v}=0$. For $v\rightarrow\infty$ it settles down to an event horizon, which is situated along $u=1.06$. It surrounds the central spacelike singularity located along the line $r=0$, which becomes singular at $u=1.38$ (the peak on the diagram).\\
The dynamical Reissner-Nordstr\"{o}m spacetime emerging from the collapse of an electrically charged scalar field ($EM$-evolution) \cite{ore03} for $\tpe_s=0.6$ is shown in Fig.\ref{fig:RNS}b. There are two horizons in the spacetime. 
The outer apparent horizon, $r_{,v}=0$, after the dynamical part of evolution, when 
$v\rightarrow\infty$, coincides with an event horizon located along $u=0.84$. 
The inner horizon, namely the Cauchy horizon is located at future null infinity, i.e., at $v\rightarrow\infty$. The 
line $r=0$ becomes central spacelike singularity at $u=3.04$.

\begin{table}
\caption{Initial profiles of field functions.}
\begin{ruledtabular}
\begin{tabular}{c|cp{5cm}|}
Family & Profile\\
\hline
&\\
$\left(f_D\right)$ & $\tpe_k\cdot v^2\cdot e^{-\left(\frac{v-c_1}{c_2}\right)^2}$\\
&\\
$\left(f_S\right)$ & $\tpe_s\cdot \sin^2\left(\pi\frac{v}{v_f}\right)
\cdot\Bigg(\cos\left(\pi\frac{2v}{v_f}\right)+i\cos\left(\pi\frac{2v}{v_f}+\delta\right)\Bigg)$\\
&\\
\end{tabular}
\end{ruledtabular}
\label{tab:initial-profiles}
\end{table}

%%%%%%%%%%%%%%%%%%%%%%%%%%%%%%%%%%%%%%%%%%%%%%%%%%%%%%%%%%%%%%%%%%%%%%%%%%%%%%%%%%%%%%%%%%%%%%%%%%%%%%%%%%%%%%%%%%%%%%%%%%%%%%%
%%%%%%%%%%%%%%%%%%%%%%%%%%%%%%%%%%%%%%%%%%%%%%%%%%%%%%%%%%%%%%%%%%%%%%%%%%%%%%%%%%%%%%%%%%%%%%%%%%%%%%%%%%%%%%%%%%%%%%%%%%%%%%%
\section{Results}
As was mentioned in the introduction, the main aim of our research was to investigate the results of 
the dynamical collapse of an electrically charged scalar field in the presence of dilaton field, while one or 
both types of fields are {\it phantom}. All the possible configurations were presented in 
Table \ref{tab:evolutions-characteristics}. 
Because of the fact that the structure of spacetimes emerging from the dynamical collapse depends on the 
unknown coupling constant $\alpha$ \cite{bor11}, we take into account $\alpha = -1$ and $\alpha = 0$. 
These values are particularly interesting because they refer to analytical models discussed in the literature. 
Specifically, $\alpha=-1$ refers to the low-energy string theory, while when $\alpha$ vanishes the dynamical collapse
of an electrically charged scalar field in presence of an uncoupled dilaton field takes place.
On the contrary, the electric coupling constant
does not exert any influence on the evolution, so we assume that it is equal to $e=0.5$.
The results of the numerical computations are depicted on Penrose diagrams, i.e., we plot lines of constant  
$r\left(u,v\right)$ in the $\left(v,u\right)$-plane.               
\par
The types of the initial field profiles, which were used in our numerical computations are presented in 
Table \ref{tab:initial-profiles}. The $\left(f_D\right)$-family refers to dilaton field, while 
the family $\left(f_S\right)$ is connected with electrically charged scalar field. The free family 
parameter $\tpe$ denotes an amplitude of the particular field. 
Namely, $\tpe_s$ is responsible for the electrically charged scalar field, while $\tpe_k$ for the dilaton field.
\par
At the beginning we will consider the separate evolutions of an electrically charged scalar field with {\it phantom} coupling of Maxwell field to gravity ($E\overline{M}$-collapse) and {\it phantom} dilaton field ($E\overline{D}$-collapse). 
Then we shall pay attention to the evolutions when more than one field is involved ({\it phantom}-$EMD$-collapses). In these situations the one-parameter cases will be considered at first. It means that the amplitudes of both collapsing fields will be taken to be equal and their common value will be denoted by $\tpe$. Afterwards, in order to complete the physical picture of the results of the examined process, for each of {\it phantom}-$EMD$ 
evolutions we shall discuss the behavior of the system in the case when the amplitudes of the considered fields 
differ. To be precise, we describe the changes in the 
spacetime structures appearing when one of the fields has a constant 
amplitude and the amplitude of the other field varies. The constant amplitudes of the electrically charged scalar 
field and dilaton field were chosen to be equal to $\tpe_s=0.6$ and $\tpe_k=0.1$, respectively. Such choice 
was motivated by the fact that these evolutions in the 
non-{\it phantom} case lead to the dynamical Reissner-Nordstr\"{o}m and Schwarzschild spacetimes as was 
described above. At the same time in the {\it phantom} cases, as will be explained in sections \ref{sec:E-M} 
and \ref{sec:E-D}, they lead to Schwarzschild and non-singular spacetimes, which are typical for 
particular {\it phantom} evolutions. 
In view of the bewildering complexity of the considered dynamical collapse it will
enable us to generalize about the role of the accompanying field (which amplitude varies) in the 
phenomena under consideration.

%%%%%%%%%%%%%%%%%%%%%%%%%%%%%%%%%%%%%%%%%%%%%%%%%%%%%%%%%%%%%%%%%%%%%%%%%%%%%%%%%%%%%%%%%%%%%%
%%%%%%%%%%%%%%%%%%%%%%%%%%%%%%%%%%%%%%%%%%%%%%%%%%%%%%%%%%%%%%%%%%%%%%%%%%%%%%%%%%%%%%%%%%%%%%
\subsection{Einstein-{\it phantom Maxwell} collapse}
\label{sec:E-M}
\begin{figure}[p]
\includegraphics[scale=0.5]{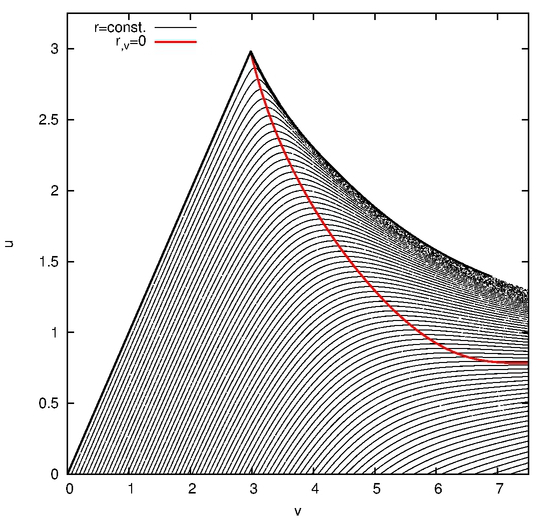}
\caption{(color online).
Lines of constant $r$ in the $(vu)$-plane for $E\overline{M}$-evolution 
with the electric coupling constant $e=0.5$. The family parameter is set to be $\tpe_s=0.6$.}
\label{fig:E-M}
\end{figure}

To begin with, we shall consider the first of the simplified versions of the theory under consideration. 
Namely, we will discuss spacetime structures emerging in the process of the dynamical collapse of the charged scalar field 
in the case of Einstein-{\it phantom Maxwell} theory ($E\overline{M}$-evolution). The representative structure 
is shown in Fig.\ref{fig:E-M}. 
The parameter characterizing the initial amplitude of the considered field was chosen as $\tpe_s=0.6$. 
It turned out that the most striking feature of the resulting
spacetime is the fact that despite the presence of an electric charge only one apparent horizon is observed.
The outcome of the considered dynamical collapse is similar to the one in the case of Schwarzschild spacetime, 
which results from the evolution of a self-gravitating electrically neutral scalar field (Fig.\ref{fig:RNS}a).
The single apparent horizon  $r_{,v}=0$ surrounds the spacelike central singularity and
settles along an event horizon for $v\rightarrow\infty$.\\
The event horizon is a null hypersurface situated along $u=0.78$. 
In comparison with the case of the dynamical collapse of the electrically charged scalar field in non-{\it phantom}
Einstein-Maxwell case ($EM$-evolution), observed for the same parameter $\tpe_s=0.6$ and depicted in Fig.\ref{fig:RNS}b,
the emerging event horizon is situated along the smaller value of retarded time. As far as the point, where the line $r=0$ becomes singular is concerned (peak on the Penrose diagram), one can
draw the same conclusion. Namely, in {\it phantom} dynamical collapse it corresponds to $u=2.98$, whereas 
in non-{\it phantom} Einstein-Maxwell evolution the singularity arises slightly later as was described above.
\par
To conclude, one can observe that in the case of Einstein-{\it phantom Maxwell} dynamical collapse the structure
of the issuing spacetime is Schwarzschild-like and the black hole forms earlier in terms of the retarded time
comparing to the non-{\it phantom} Einstein-Maxwell evolution. The comparison between $E\overline{M}$ and $EM$ cases 
allows us to draw an interesting conclusion concerning the correlation between the attractive/repulsive nature of forces 
and spacetime structures. As was already mentioned, gravitational force is always attractive and supports the collapse. 
The electrostatic force in $EM$-evolution is repulsive and the Reissner-Nordstr\"{o}m spacetime is the outcome 
of some kind of a compromise between the two forces. In contrast, the electrostatic force in $E\overline{M}$-evolution 
is attractive just as gravitational force. In this case a simpler, 
Schwarzschild-like spacetime emerges.

%%%%%%%%%%%%%%%%%%%%%%%%%%%%%%%%%%%%%%%%%%%%%%%%%%%% 
%%%%%%%%%%%%%%%%%%%%%%%%%%%%%%%%%%%%%%%%%%%%%%%%%%%%%%%%%%%%%%%%%%%%%%%%%%%%%%%%%%%%%%%%%%%%%%%%%%%%%%%%%%%
\subsection{Einstein-{\it phantom dilaton} collapse}
\label{sec:E-D}
\begin{figure}[p]
\includegraphics[scale=0.5]{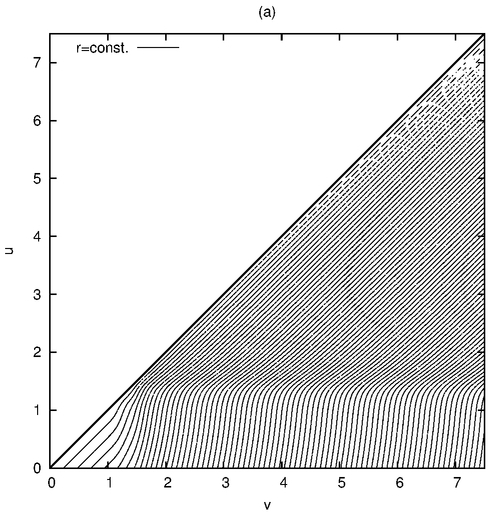}
\includegraphics[scale=0.5]{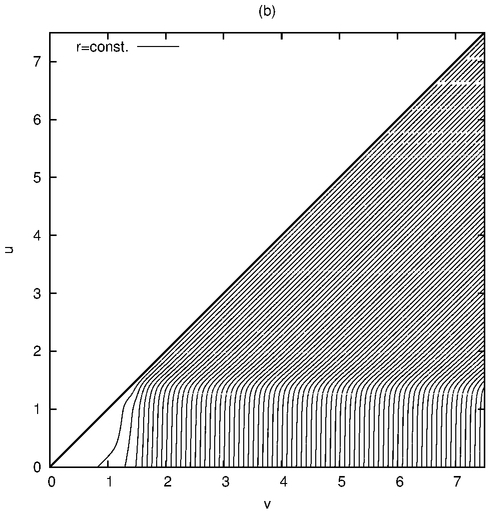}
\includegraphics[scale=0.5]{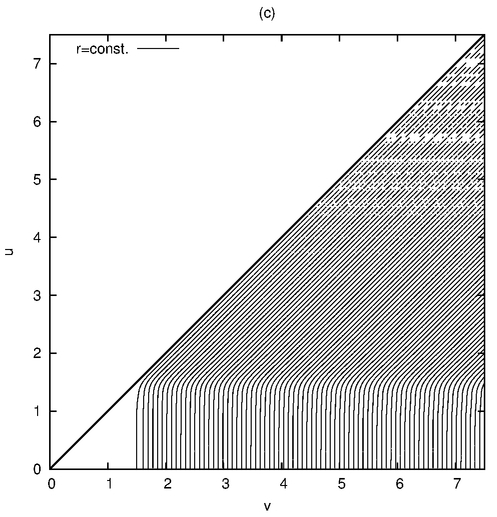}
\caption{(color online).
Lines of constant $r$ in the $(vu)$-plane for $E\overline{D}$-evolution 
with the family parameter equal to (a) $\tpe_k=0.25$, (b) $\tpe_k=0.5$ and (c) $\tpe_k=1$.}
\label{fig:E-D}
\end{figure}

To proceed further, let us consider the second simplification, i.e., dynamical collapse in the case
of Einstein-{\it phantom dilaton} system
({$E\overline{D}$}-evolution), where dilaton field evolves under the influence of gravitational field. 
This case simply refers to the evolution of {\it phantom} electrically neutral scalar field. 
A vast range of the parameters $\tpe_k$ characterizing the initial amplitude of the field was examined.
The results of the numerical calculations for $\tpe_k$ equal to $0.25$, $0.5$ and $1$ are shown in Fig.\ref{fig:E-D}. 
An important notion concerning the evolution in question is the fact that 
for every considered dilaton field amplitude the emerging spacetime is non-singular, that is black holes do not form. 
The only influence of the parameter $\tpe_k$ on the resulting
spacetime structure is the fact that 
the bigger value of $\tpe_k$ we take the more significantly $r$-coordinate varies in the vicinity of the center
of spacetime (pushing out the lines of $r=const.$ in the vicinity of the point $u=v=0$). 
The aforementioned behaviour is in complete contrast to the dynamical evolution in Einstein-scalar theory 
($ED$-evolution). In this case the dynamical collapse of non-{\it phantom}
scalar field, for appropriately large values of the parameter $\tpe_k$ results in a singular,
Schwarzschild-type spacetime, as was shown in Fig.\ref{fig:RNS}a.
\par
Summing it all up, the influence of the {\it phantom} nature of the scalar field is even more 
extreme in the electrically neutral case 
than in the charged one. In Einstein-{\it phantom dilaton} theory black holes do not appear in comparison with 
Einstein-scalar case, where Schwarzschild-type black holes are obtained. It may be also explained in terms of 
the nature of acting forces. Attractive gravitational force cooperates with an attractive force during $ED$-collapse 
and competes with repulsive dilatonic force in $E\overline{D}$-evolution. It turns out that in the latter case 
gravity is not able to overcome the repulsive nature of {\it phantom} field.

%%%%%%%%%%%%%%%%%%%%%%%%%%%%%%%%%%%%%%%%%%%%%%%%%%%%%%%%%%%%%%%%%%%%%%%%%%%%%%%%%%%%%%%%%%%%%%%%%%%%%%%%%%%%%%%%%%%% 
%%%%%%%%%%%%%%%%%%%%%%%%%%%%%%%%%%%%%%%%%%%%%%%%%%%%%%%%%%%%%%%%%%%%%%%%%%%%%%%%%%%%%%%%%%%%%%%%%%%%%%%%%%%%%%%%%%% 
\subsection{Einstein-{\it phantom Maxwell}-dilaton collapse}
\begin{figure}[p]
\includegraphics[scale=0.5]{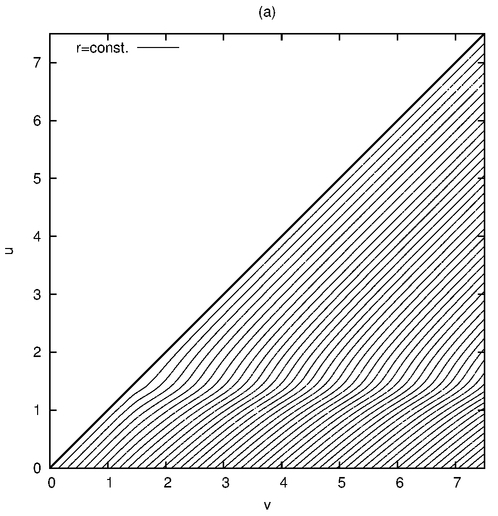}
\includegraphics[scale=0.5]{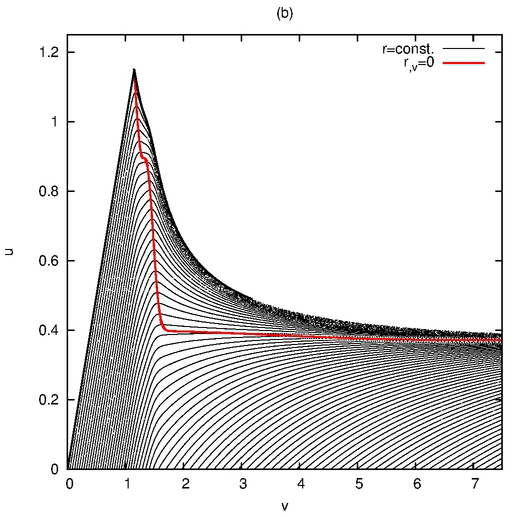}
\caption{(color online).
Lines of constant $r$ in the the $(vu)$-plane for $E\overline{M}D$-evolution 
with the electric coupling constant $e=0.5$ 
and the dilatonic coupling constant $\alpha$ equal to $-1$ and $0$.
The family parameter is taken to be (a) $\tpe=0.06$ and (b) $\tpe=0.175$.}
\label{fig:E-MD-10}
\end{figure}

The first theory concerning the collapse of both considered fields, namely the dilaton and electrically charged 
scalar ones, is Einstein-{\it phantom Maxwell}-dilaton theory ($E\overline{M}D$-evolution). It should be emphasized that there are no differences between spacetime structures resulting from the process for the coupling constant $\alpha$ equal to $-1$ and $0$ while values of field amplitudes do not differ. In Fig.\ref{fig:E-MD-10} the outcome of an evolution in $E\overline{M}D$-theory for both considered values of the dilatonic coupling constant is presented. The value of parameter $\tpe$ was taken to be equal to $0.06$ and $0.175$.
For small values of $\tpe$ the spacetime is non-singular. For its bigger values the Schwarzschild-like black hole forms. We achieve a single apparent horizon $r_{,v}=0$ surrounding spacelike singularity. When $v$-coordinate tends to infinity the apparent horizon coincides with the event horizon of a newly born black hole.
\par
This situation is also in agreement with the previous analyzes of the nature of acting forces. In this case 
all of them are attractive, hence for the sufficiently large value of the amplitudes the simplest singular, Schwarzschild-type structure forms.
\par
\begin{figure}[p]
\includegraphics[scale=0.5]{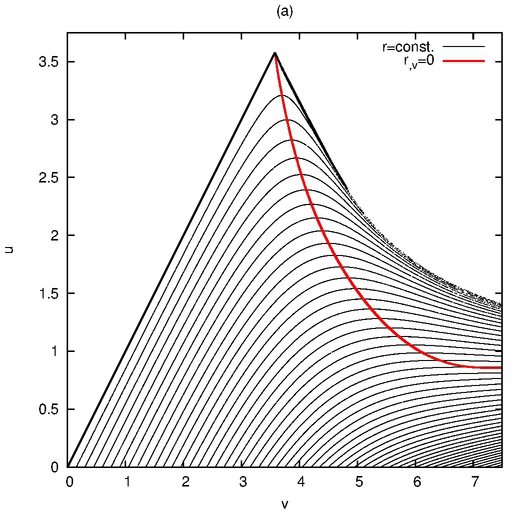}
\includegraphics[scale=0.5]{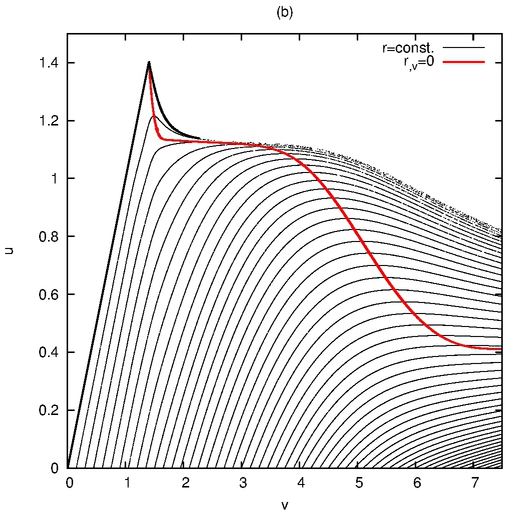}
\caption{(color online).
Lines of constant $r$ in $(vu)$-plane for the $E\overline{M}D$ evolution with a varying dilaton field amplitude. 
The coupling 
constants have the same values as in Fig.\ref{fig:E-MD-10}.
The family parameter for the electrically charged scalar field $\tpe_s=0.6$, 
while for the dilaton field (a) $\tpe_k=0.01$ and (b) $\tpe_k=0.09$.}
\label{fig:E-MDaD-10}
\end{figure}
\begin{figure}[p]
\includegraphics[scale=0.5]{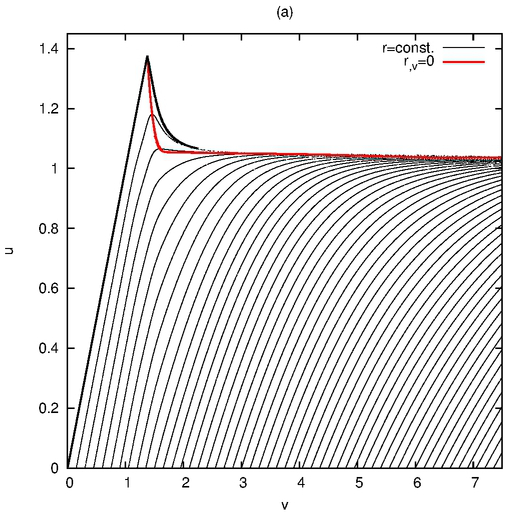}
\includegraphics[scale=0.5]{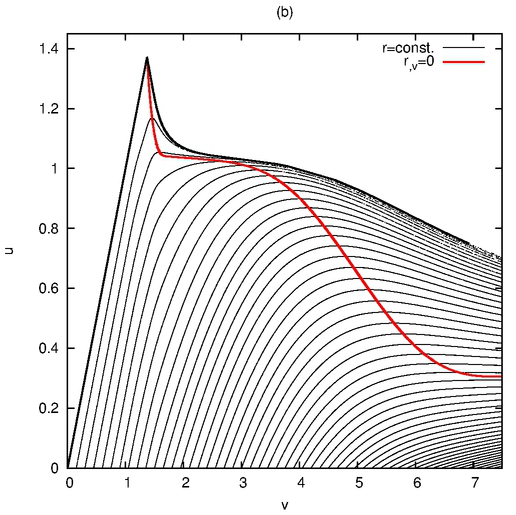}
\caption{(color online).
Lines of constant $r$ in the $(vu)$-plane for the $E\overline{M}D$ evolution with varying 
electrically charged scalar field amplitude. The coupling constants have the same values as in 
Fig.\ref{fig:E-MD-10}. The family parameter for the dilaton field is equal to $\tpe_k=0.1$, while for the electrically 
charged scalar field field is set to be (a) $\tpe_s=0.35$ and (b) $\tpe_s=0.6$, respectively.}
\label{fig:E-MDaM-10}
\end{figure}

In order to proceed to considerations of how various fields influence on the dynamical collapse
in question, we kept an amplitude of electrically charged scalar field $\tpe_s$ constant and equal to $0.6$ 
and varied the amplitude of dilaton field. The structures of spacetimes for $\alpha=-1,0$ and $\tpe_k$ equal to $0.01$ and $0.09$ are shown in Fig.\ref{fig:E-MDaD-10}.
Then, we performed computations the other way round. Namely, we changed an amplitude of electrically charged scalar field $\tpe_s$ while keeping the other amplitude $\tpe_k$ constant and equal to $0.1$. In Fig.\ref{fig:E-MDaM-10} we depicted spacetime structures for the amplitude of electrically charged scalar field equal to $0.35$ and $0.6$ for both considered values of dilatonic coupling constant. In all the cases we obtained dynamic Schwarzschild spacetimes.
\par
For small values of a varying amplitude we observed Schwarzschild-like spacetimes like the one shown in Fig.\ref{fig:RNS}a, that is a central spacelike singularity surrounded by one apparent horizon $r_{,v}=0$, 
which after a dynamical part of evolution settles along an event horizon for $v\rightarrow\infty$. 
For larger values of a varying amplitude slightly different behaviour was revealed, namely two 
dynamical stages during the evolution were observed. After achieving a constant value of $u$-coordinate 
an apparent horizon $r_{,v}=0$ changes its position to another value of retarded time, which is smaller than 
the previous one. As $v\rightarrow\infty$ the horizon again settles along $u=const.$ indicating the position of an event horizon of a final black hole. Of course a spacelike singularity is situated in the center of it. Such structure also corresponds to Schwarzschild-type.\\
An increase of $\tpe_k$ results in earlier (in terms of retarded time) appearance of a point, where the line $r=0$ becomes singular (Fig.\ref{fig:E-MDaD-10}). This increase is also connected with the fact that an event horizon is situated along a smaller constant value of $u$-coordinate. The decrease of a value of retarded time corresponding to a location of an event horizon is more significant than the decrease of $u$ referring to the first singular point at $r=0$.
Within the set of solutions with altering amplitude of {\it phantom} counterpart of electrically charged scalar field (Fig.\ref{fig:E-MDaM-10}), the changes of $\tpe_s$ do not affect the location of a point, where the line $r=0$ becomes singular. At the same time the location of the event horizon moves towards lower values of $u$-coordinate when $\tpe_s$ increases.
\par
The energy momentum tensor (\ref{ten})--(\ref{ten-2}) for $E\overline{M}D$-theory in the case of $\alpha = -1$ has the following form:
\ben
T_{\mu \nu}(F, \phi) &=& - e^{- 2 \phi} T_{\mu \nu}(F) + T_{\mu \nu}(\phi), \\ \nonumber
T_{\mu \nu}(\phi, F, \psi, \zpsi, A) &=& \tT_{\mu \nu}(\psi, \zpsi, A) + T_{\mu \nu}(F, \phi),
\een
while for $\alpha = 0$:
\ben
T_{\mu \nu}(F, \phi) &=& - T_{\mu \nu}(F) + T_{\mu \nu}(\phi), \\ \nonumber
T_{\mu \nu}(\phi, F, \psi, \zpsi, A) &=& e^{2 \phi}\tT_{\mu \nu}(\psi, \zpsi, A) + T_{\mu \nu}(F, \phi),
\een
where $\tT_{\mu \nu}(\psi, \zpsi, A)$, $T_{\mu \nu}(F)$ and $T_{\mu \nu}(\phi)$ denote the parts connected with electrically charged, Maxwell and dilaton fields, respectively. 
The coexisting effects of the dilaton field and {\it phantom} counterpart of an electrically charged scalar 
field result in an appearance of Schwarzschild black hole. Even though the changes of the amplitudes 
of particular fields change their relative impact on the collapse, its outcome is in all cases similar. 
\par
It is worth mentioning that analytic studies of singular static configurations in $E\overline{M}D$-theory were conducted in Refs.\cite{gib96, cle09}. It turned out that in the considered case the solutions are Schwarzschild-type, i.e., the spacetimes consist of one horizon surrounding central singularity. This conclusion was confirmed by our computations also in the case of dynamical collapse.

%%%%%%%%%%%%%%%%%%%%%%%%%%%%%%%%%%%%%%%%%%%%%%%%%%%%%%%%%%%%%%%%%%%%%%%%%%%%%%%%%%%%%%%%%%%%%%%%
%%%%%%%%%%%%%%%%%%%%%%%%%%%%%%%%%%%%%%%%%%%%%%%%%%%%%%%%%%%%%%%%%%%%%%%%%%%%%%%%%%%%%%%%%%%%%%%%
\subsection{Einstein-Maxwell-{\it phantom dilaton} collapse}
\begin{figure}[p]
\includegraphics[scale=0.5]{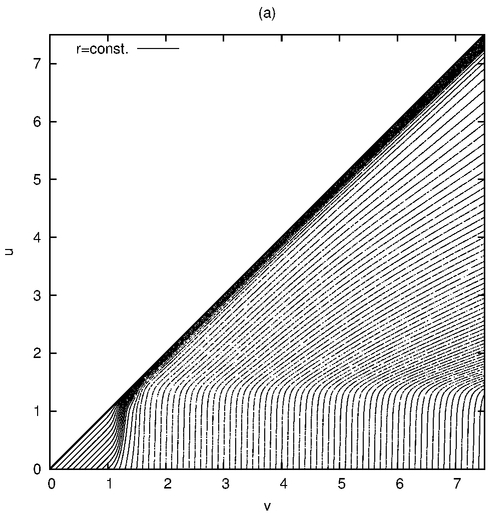}\includegraphics[scale=0.5]{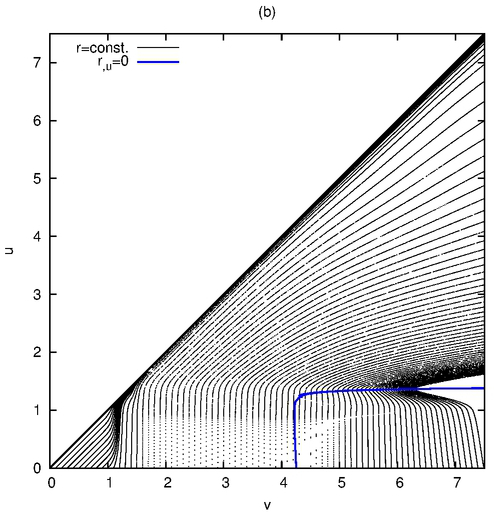}
\includegraphics[scale=0.5]{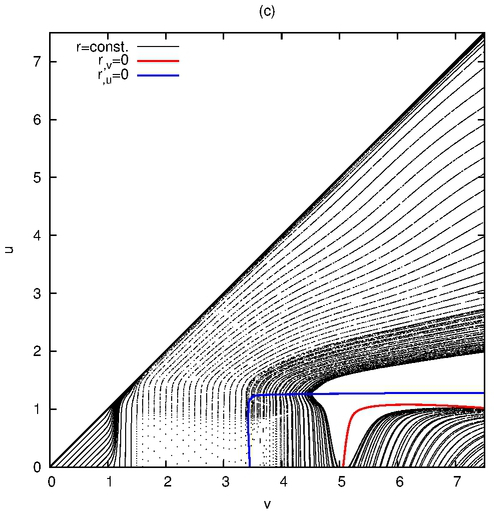}
\caption{(color online).
Lines of constant $r$ in the $(vu)$-plane for $EM\overline{D}$-evolution with 
the electric and the dilatonic coupling constants equal to $e=0.5$ and $\alpha=-1$, respectively. 
The family parameter is equal to (a) $\tpe=0.5$, (b) $\tpe=0.85$ and (c) $\tpe=1.1$.}
\label{fig:EM-D-1}
\end{figure}

\begin{figure}[p]
\includegraphics[scale=0.5]{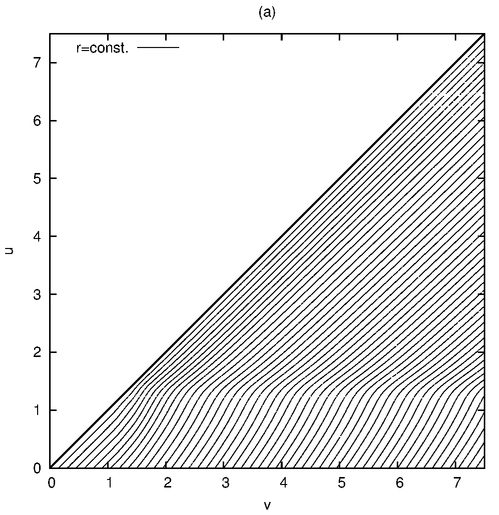}\includegraphics[scale=0.5]{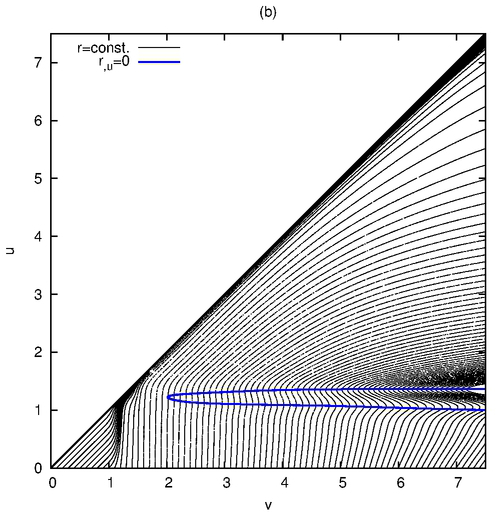}
\includegraphics[scale=0.5]{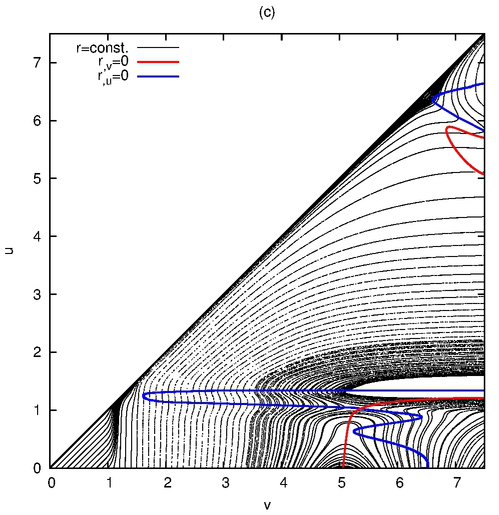}
\caption{(color online).
Lines of constant $r$ in the $(vu)$-plane for $EM\overline{D}$-evolution with 
the electric and the dilatonic coupling constants equal to $e=0.5$ and $\alpha=0$, respectively. 
The family parameter is provided by (a) $\tpe=0.1$, (b) $\tpe=0.8$ and (c) $\tpe=1.1$.}
\label{fig:EM-D0}
\end{figure}

The next type of the theory we shall be interested in is Einstein-Maxwell-{\it phantom dilaton} gravity
($EM\overline{D}$-evolution). In this theory the electrically charged scalar field coupled to {\it phantom} 
dilaton field collapses gravitationally. The plots bounded with the dynamical collapse in 
$EM\overline{D}$-evolution, for the dilatonic coupling constant equal to 
$-1$ and $0$ are shown in Figs.\ref{fig:EM-D-1} and \ref{fig:EM-D0}, respectively.
In our considerations, for the case of $\alpha = -1$, we have chosen the parameter $\tpe$ to be equal to 
$0.5$, $0.85$ and $1.1$. When the coupling constant $\alpha$ is equal to zero 
$\tpe$ equals to $0.1$, $0.8$ and $1.1$.
\par
For all values of the family parameter $\tpe$ the non-singular spacetimes emerge for both considered values of 
$\alpha$. For small values of the parameter $\tpe$ there are no horizons in spacetime. 
Refining our numerical studies to the larger values of $\tpe$, we obtain an apparent horizon
$r_{,u}=0$ for small values of retarded time. For $\alpha=-1$, it settles along constant $u$-coordinate 
for $v\rightarrow\infty$, while for $\alpha=0$ it forms an open loop with ending points at $v\to\infty$. It 
happens that for even bigger values of the amplitudes another type of an apparent horizon, 
namely $r_{,v}=0$, emerges. We observe both $r_{,u}=0$ and $r_{,v}=0$ types of the apparent horizons. 
However, their presence in spacetime is not connected with the appearance of the singularity.\\
For $\alpha=-1$ the results are depicted in Fig.\ref{fig:EM-D-1}.
One observes that the appearance of $r_{,u}=0$ and $r_{,v}=0$ apparent horizons for small values of $u$-coordinate
develops a {\it gap} between $r=const.$ lines at $v\rightarrow\infty$. The adjacent lines move away very quickly. 
As was mentioned above, for large values of parameter $\tpe$ both types of apparent horizons emerge.\\
For the coupling constant $\alpha=0$ the results are shown in Fig.\ref{fig:EM-D0}.
One gets that for large values of the parameter $\tpe$ and small values of the $u$-coordinate as well as 
for large values of both retarded and advanced times, both $r_{,u}=0$ and $r_{,v}=0$ apparent horizons 
exist in the non-singular spacetime.
\par
In the considered $EM\overline{D}$-theory the repulsive electromagnetic and dilatonic forces compete with 
the attractive gravitational force after all leading to non-singular spacetimes 
with single or multiple horizons.
%%%%%%%%%%%%%%%%%%%%%%%%%%%%%%%%%%%%%%%%%%%%%%%%%%%%%%%%%%%%%%%%%%
\par
\begin{figure}[p]
\includegraphics[scale=0.5]{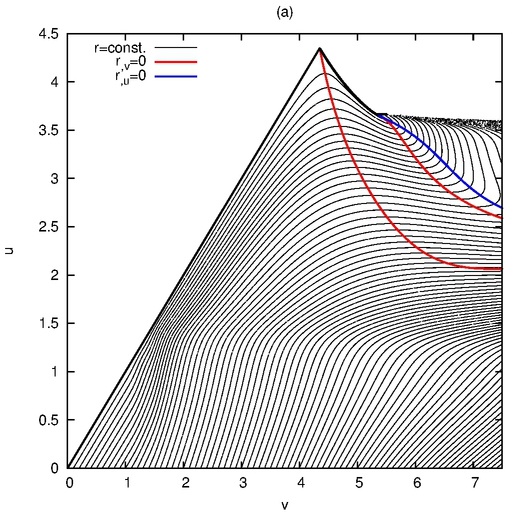}\includegraphics[scale=0.5]{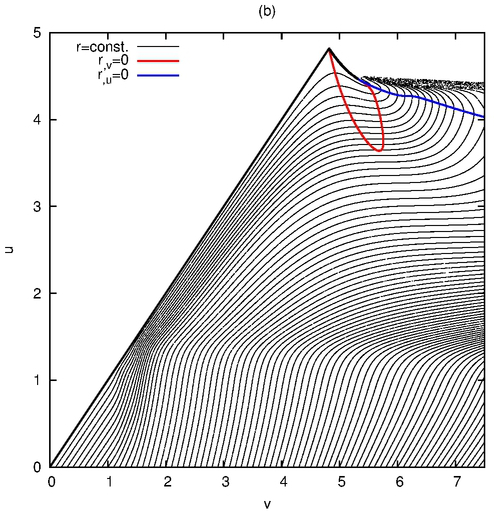}
\includegraphics[scale=0.5]{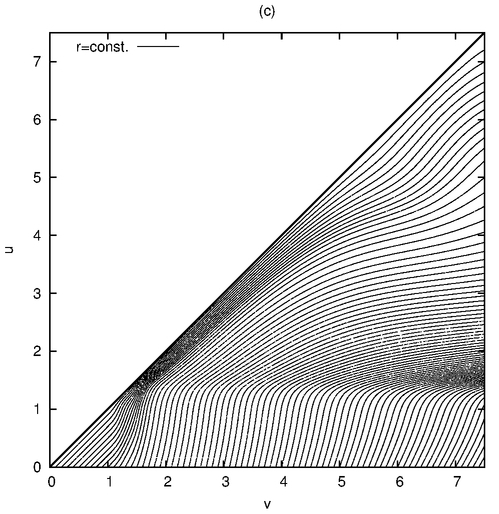}
\caption{(color online).
Lines of constant $r$ in $(vu)$-plane for $EM\overline{D}$ evolution with a varying dilaton field amplitude. 
The electric and dilatonic coupling constants are equal to $e=0.5$ and $\alpha=-1$, respectively. 
The family parameter for the electrically charged scalar field is equal to $\tpe_s=0.6$, while for the 
dilaton field is set as (a) $\tpe_k=0.15$, (b) $\tpe_k=0.2$ and (c) $\tpe_k=0.25$.}
\label{fig:EM-DaD-1}
\end{figure}

In order to get a deeper insight into the physical processes taking place during $EM\overline{D}$-collapse
we shall vary amplitudes of {\it phantom} dilaton and of electrically charged scalar field.
We commence with altering an amplitude of {\it phantom} dilaton 
while an amplitude of the other field remains constant and equal to $\tpe_s=0.6$.
For $\alpha=-1$ the structures formed when $\tpe_k$ equals to $0.15$, $0.2$ and $0.25$ are shown 
in Fig.\ref{fig:EM-DaD-1}. For small values of the parameter $\tpe_k$ we get the singular spacetime 
with two kinds of apparent horizons, but
when the amplitude of {\it phantom} dilaton grows the spacetime becomes non-singular. 
The growing amplitude of {\it phantom} dilaton field suppresses emergence of a singularity. 
This result seems to be intuitive because {\it phantom} field contributes negatively to the total energy of the system.\\
As was mentioned above, for small values of the {\it phantom} dilaton amplitude the spacetime became
singular but with two types apparent horizons $r_{,u}=0$ and $r_{,v}=0$ in the domain of integration.
For values of amplitudes not exceeding $0.15$ there are two branches of $r_{,v}=0$ horizon.
The first one settles along an event horizon for $v\rightarrow\infty$. 
With an increasing value of the {\it phantom}
dilaton field amplitude, the singularity as well as the horizon appears at increasing $u$-coordinate. The 
second branch of the horizon $r_{,v}=0$ lies inside the first one and it begins at the point, where the 
line of singular $r=0$ ends. The apparent horizon $r_{,u}=0$ is situated inside the second 
branch of $r_{,v}=0$ horizon and begins at the same point. For small values of parameter 
$\tpe_k$ these two lines almost coincide. The structure of an area beyond the horizon 
$r_{,u}=0$ is impossible to interpret fully and clearly, because calculations cannot be 
carried out until $r=0$ and they end at some positive value of $r$. Such structure with 
two branches of $r_{,v}=0$ apparent horizon and an apparent horizon $r_{,u}=0$ inside them 
may be interpreted as a dynamical wormhole \cite{hwa11}. On the other hand,
for greater
values of the parameter $\tpe_k$ (larger than $0.2$) the two branches of the horizon $r_{,v}=0$ 
form a loop with two ends coinciding with ending points of singular part of line $r=0$. The horizon 
$r_{,u}=0$ remains outside it and continuously tends towards smaller values of $u$-coordinate 
as $v\rightarrow\infty$. This structure corresponds to an unphysical 
situation of naked singularity in spacetime \cite{hwa11}. For even bigger values 
of amplitudes of dilaton field, from approximately $\tpe_k=0.25$, the spacetime is non-singular with no visible horizons.\\
A close inspection of the energy momentum tensor (\ref{ten})--(\ref{ten-2}) for $EM\overline{D}$-theory 
in the considered case of $\alpha = -1$ reveals that one gets the following:
\ben \label{tt}
T_{\mu \nu}(F, \phi) &=& e^{- 2 \phi} T_{\mu \nu}(F) - T_{\mu \nu}(\phi), \\ \nonumber
T_{\mu \nu}(\phi, F, \psi, \zpsi, A) &=& \tT_{\mu \nu}(\psi, \zpsi, A) + T_{\mu \nu}(F, \phi).
\een
It can be seen that the factor $e^{- 2 \phi}$ diminishes the influence of the Maxwell field on the collapse, i.e.,
the bigger amplitude of {\it phantom} dilaton field we consider the smaller influence is of the $U(1)$-gauge field
we obtain. When $\tpe_k$ increases the strong repulsive effect of the {\it phantom} dilaton field begins to play the dominant role leading to non-singular spacetimes.
\par
\begin{figure}[p]
\includegraphics[scale=0.5]{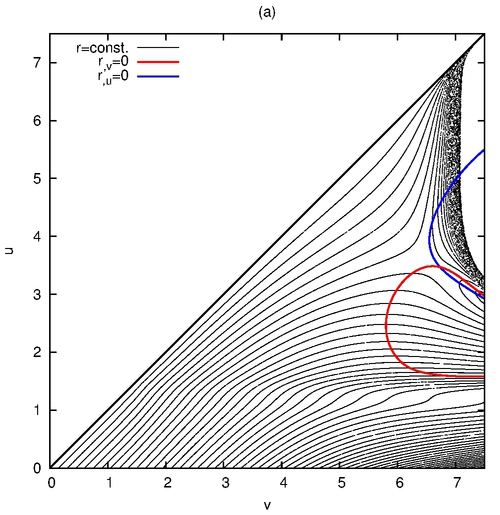}\\
\includegraphics[scale=0.5]{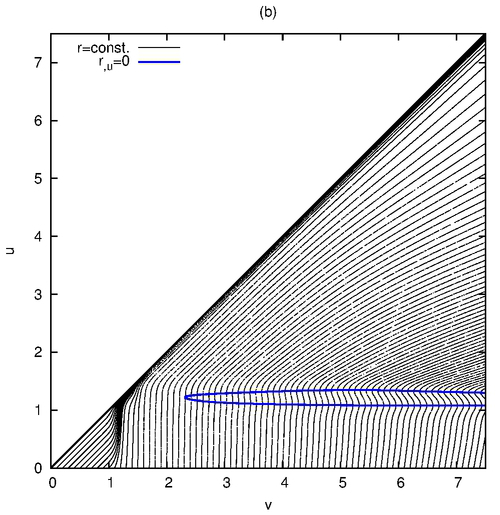}
\caption{(color online).
Lines of constant $r$ in the $(vu)$-plane for the evolution described in Fig.\ref{fig:EM-DaD-1} with the 
dilatonic coupling constant equal to $\alpha=0$ and the family parameter for dilaton 
field given by (a) $\tpe_k=0.05$ and (b) $\tpe_k=0.8$.}
\label{fig:EM-DaD0}
\end{figure}

Different situation takes place for Einstein-Maxwell-{\it phantom dilaton} collapse without 
dilaton coupling, that is for $\alpha=0$. The structures of the emerging 
spacetimes for $\tpe_k$ equal 
to $0.05$ and $0.8$ are shown in Fig.\ref{fig:EM-DaD0}. The spacetimes are non-singular for all 
values of the parameter $\tpe_k$. It means that the presence of {\it phantom}
dilaton field, not interacting with the rest of matter in spacetime, prevents black hole formation.  
The spacetime is non-singular with both types of horizons at first and then, for $\tpe_k$ exceeding 
$0.7$, only one open loop-shaped
apparent horizon $r_{,u}=0$ is observable for the small values of the $u$-coordinate.\\
On the other hand, the energy momentum tensor for $\alpha = 0$ is provided by
\ben \label{ta0}
T_{\mu \nu}(F, \phi) &=& T_{\mu \nu}(F) - T_{\mu \nu}(\phi), \\ \nonumber
T_{\mu \nu}(\phi, F, \psi, \zpsi, A) &=& e^{2 \phi} \tT_{\mu \nu}(\psi, \zpsi, A) + T_{\mu \nu}(F, \phi).
\een
Eq.(\ref{ta0}) reveals that the growth of the {\it phantom} dilaton field amplitude plays the dominant role
while the factor $e^{2 \phi}$ also enlarges the influence of the 
electrically charged scalar field. Therefore due to these facts 
the domination of the {\it phantom} dilaton field seems to be a key factor in the process of the dynamical collapse
for this case.
\par
\begin{figure}[p]
\includegraphics[scale=0.5]{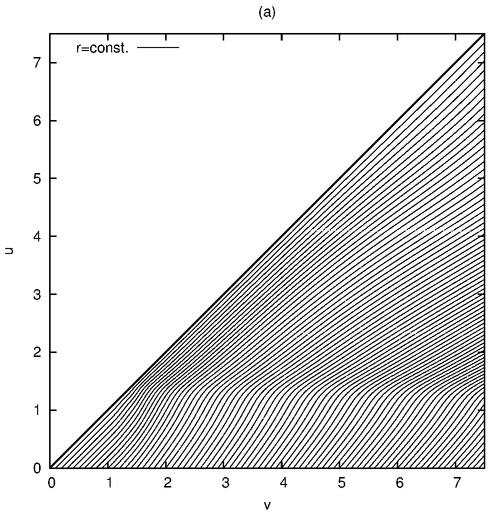}
\includegraphics[scale=0.5]{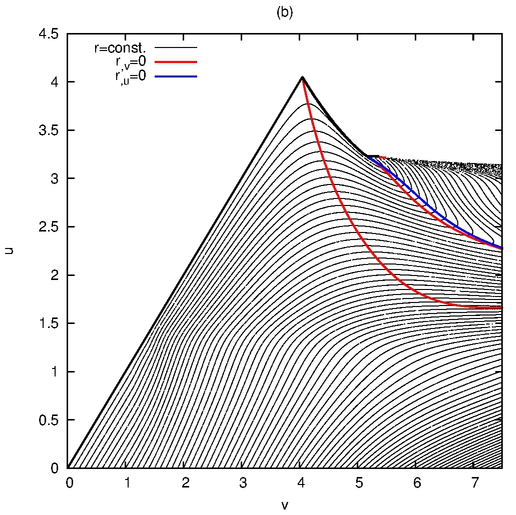}
\caption{(color online).
Lines of constant $r$ in the $(vu)$-plane for $EM\overline{D}$ evolution with a varying electrically 
charged scalar field amplitude. The electric and dilatonic coupling constants are equal to $e=0.5$ and $\alpha=-1$, 
respectively. The family parameter for the dilaton field is $\tpe_k=0.1$, while for the 
electrically charged scalar field field is chosen to be (a) $\tpe_s=0.3$ and (b) $\tpe_s=0.6$.}
\label{fig:EM-DaM-1}
\end{figure}

The next set of solutions referring to Einstein-Maxwell-{\it phantom dilaton} theory involves 
spacetimes emerging from a collapse when an amplitude of {\it phantom} dilaton field is constant and 
an amplitude of electrically charged scalar field varies. The value of the constant amplitude was set as $\tpe_k=0.1$. 
In Fig.\ref{fig:EM-DaM-1} the structures of spacetimes emerging from the collapse for $\alpha=-1$ were depicted. The 
amplitudes of the scalar field $\tpe_s$ were chosen to be equal to $0.3$ and $0.6$, respectively. 
One can observe that for small values of the electrically charged scalar field amplitude
the repulsive character of {\it phantom} dilaton field combined with small amount of the electrically charged 
scalar field results
in the nonsingularity of the emerging spacetime.
On the other hand, when the role of {\it phantom} dilaton field is diminished, i.e., the 
amplitude of electrically charged scalar field grows, one 
attains the singular spacetime with two branches of $r_{,v}=0$ apparent horizon -- the outer and the inner one 
as well as an apparent horizon $r_{,u}=0$ located within them. One remarks that a dynamical wormhole structure appears. 
Namely, the emerging black hole gains {\it phantom} dilaton matter violating 
the null energy condition and this phenomenon causes formation of a wormhole.\\
As may be inferred from Eq.(\ref{tt}) the growth of an electrically charged scalar field amplitude 
is able to neutralize the diminishing effect of the factor $e^{-2 \phi}$ coexisting 
with the repulsive tendency of {\it phantom} dilaton field. It finally leads to the formation of a dynamic wormhole.
\par
\begin{figure}[p]
\includegraphics[scale=0.5]{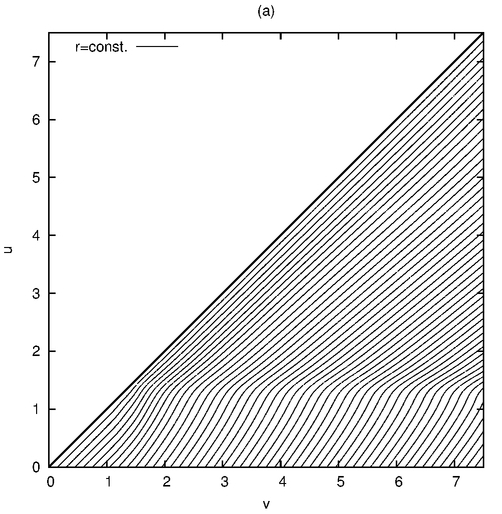}\\
\includegraphics[scale=0.5]{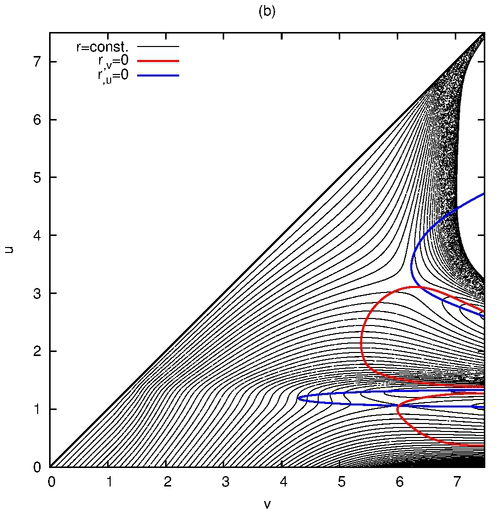}
\caption{(color online).
Lines of constant $r$ in the $(vu)$-plane for the evolution described in Fig.\ref{fig:EM-DaM-1} with 
the dilatonic coupling constant equal to $\alpha=0$ and the family parameter for electrically charged 
scalar field field (a) $\tpe_s=0.2$ and (b) $\tpe_s=0.7$.}
\label{fig:EM-DaM0}
\end{figure}

The structures of spacetimes emerging from $EM\overline{D}$-evolution with constant value of an 
amplitude $\tpe_k$ and dilatonic coupling constant $\alpha=0$ are shown in Fig.\ref{fig:EM-DaM0}. The amplitude of charged scalar 
field $\tpe_s$ was set as equal to $0.2$ and $0.7$. The spacetimes for all values of a varying amplitude are 
non-singular. For small values of it there are no horizons visible in the domain of integration. In turn, for 
bigger 
values of $\tpe_s$ apparent horizons $r_{,v}=0$ and $r_{,u}=0$ appear in spacetime, but none of them 
reach the line $r=0$, which for this reason does not become singular. 
The emergence of the apparent horizons
is accompanied by squeezing the lines of $r=const.$ in the vicinity of the line $r=0$, in the area lying above 
the $r_{,v}=0$ horizon in terms of the $u$-coordinate.\\
From Eq.(\ref{ta0}) we can observe that the factor containing {\it phantom} dilaton field $e^{2 \phi}$ plays 
an important role together with the growth of the electrically charged scalar field amplitude
in the domain of integration. This is closely related to the fact that {\it phantom} dilaton field 
suppresses appearance of black hole, while an increasing amplitude of the electrically charged scalar 
field relatively to dilaton field amplitude forces the formation of horizons in the underlying spacetime.

%%%%%%%%%%%%%%%%%%%%%%%%%%%%%%%%%%%%%%%%%%%%%%%%%%%%%%%%%%%%%%%%%%%%%%%%%%%%%%%%%%%%%%%%%%%%%%%%%%%%%%%%%%%
%%%%%%%%%%%%%%%%%%%%%%%%%%%%%%%%%%%%%%%%%%%%%%%%%%%% 
\subsection{Einstein-{\it phantom Maxwell}-{\it phantom dilaton} collapse}
\begin{figure}[p]
\includegraphics[scale=0.5]{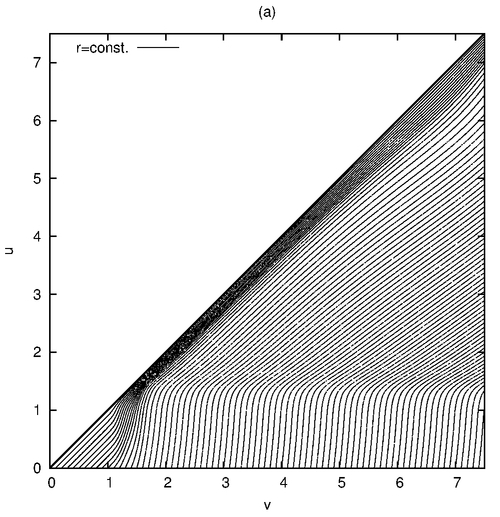}\includegraphics[scale=0.5]{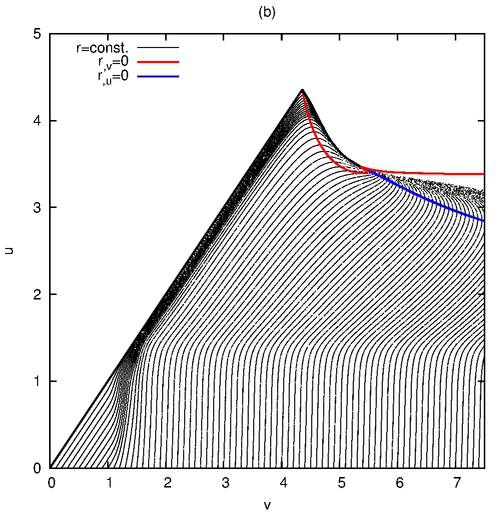}
\includegraphics[scale=0.5]{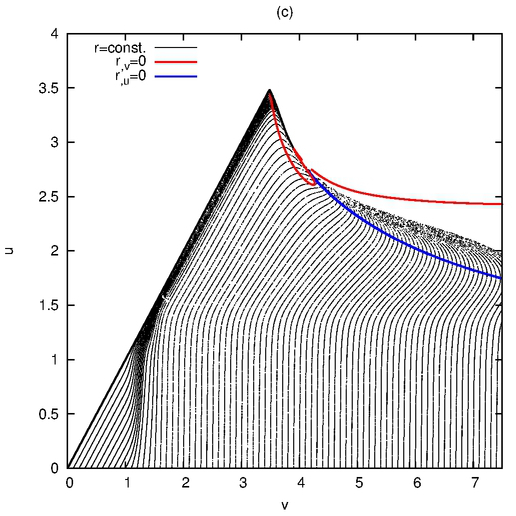}
\caption{(color online).
Lines of constant $r$ in the $(vu)$-plane for $E\overline{MD}$-evolution with the electric and 
dilatonic coupling constants equal to $e=0.5$ and $\alpha=-1$, respectively. The family 
parameter is set (a) $\tpe=0.3$, (b) $\tpe=0.4$ and (c) $\tpe=0.5$.}
\label{fig:E-M-D-1}
\end{figure}

\begin{figure}[p]
\includegraphics[scale=0.5]{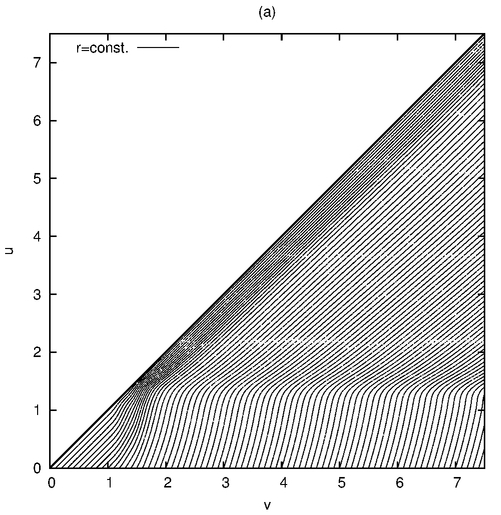}
\includegraphics[scale=0.5]{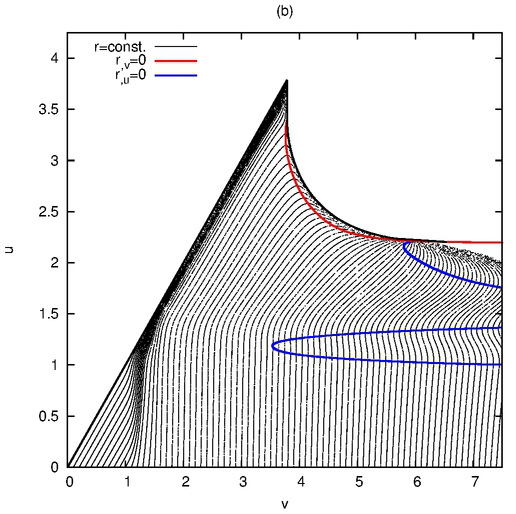}
\caption{(color online).
Lines of constant $r$ in the $(vu)$-plane for $E\overline{MD}$-evolution with the 
electric and dilatonic coupling constants equal to $e=0.5$ and $\alpha=0$, respectively. The family p
arameter is given by (a) $\tpe=0.2$ and (b) $\tpe=0.5$.}
\label{fig:E-M-D0}
\end{figure}

The last examined case is Einstein-{\it phantom Maxwell}-{\it phantom dilaton} 
theory ($E\overline{MD}$-evolution).
The resulting spacetimes formed during the process of the dynamical collapse are depicted in 
Figs.\ref{fig:E-M-D-1} and \ref{fig:E-M-D0}. The plots were made for the dilaton coupling 
constant equal to $-1$ and $0$, respectively. 
For $\alpha=-1$ the amplitudes of the considered fields were chosen to be equal to $0.3$, $0.4$ and $0.5$. 
In the case of $\alpha=0$, the parameter $\tpe$ was equal to $0.2$ and $0.5$.
\par
For both values of $\alpha$ the spacetimes for small values of $\tpe$ are non-singular without any horizons.
Similarly to the previously analyzed dynamical collapse, in the majority of cases both types of 
apparent horizons, namely $r_{,v}=0$ and  $r_{,u}=0$
are visible in the domain of integration. Moreover, the appearance of the apparent horizon $r_{,v}=0$ 
for larger values of the free family parameter always ensues from the formation of singularity in the considered spacetime. What is more, the bigger value
of the parameter one considers the earlier in terms of retarded time the singularity at $r=0$ appears.\\
In Fig.\ref{fig:E-M-D-1}, plotted for the coupling constant $\alpha=-1$, 
one observes a loop-shaped part of the apparent horizon $r_{,v}=0$ near the peak of the line $r=0$, where it 
becomes singular.
The size of a loop-shaped apparent horizon decreases with increasing field parameter $\tpe$.
What is more, the loop surrounds the singular part of the $r=0$ line. 
Outside the loop the singularity is absent or impossible to localize.
When $v$-coordinate tends to infinity, the apparent horizon $r_{,v}=0$ settles 
along constant $u$-coordinate.
Moreover, the other apparent horizon begins at the point, where the singular $r=0$ 
ends and tends towards smaller values of retarded time as $v\to\infty$. As was already 
stated in the previous section, such structure corresponds to naked singularity in spacetime.\\
In Fig.\ref{fig:E-M-D0} we presented the results for the uncoupled case, that is for $\alpha = 0$.
One notices that for the large values of parameter $\tpe$ the spacetime 
relaxes to the one containing naked singularity. There is a loop-shaped 
apparent horizon surrounding 
the central spacelike singularity and it tends towards $u=const.$ as $v\rightarrow\infty$.
Additionally, there are two branches of the horizon $r_{,u}=0$. One of them is connected 
with the naked singularity. The other one forms an open loop similar to the ones observed 
in spacetimes emerging from the $EM\overline{D}$-evolutions conducted for $\alpha=0$.
\par
In conclusion it may be stated that spacetimes emerging from the collapse 
in Einstein-{\it phantom Maxwell}-{\it phantom dilaton} theory with equal amplitudes of both fields 
are either non-singular or they contain naked singularities. In this case 
only the dilatonic force is repulsive, the other two forces have attractive character.
\par
\begin{figure}[p]
\includegraphics[scale=0.5]{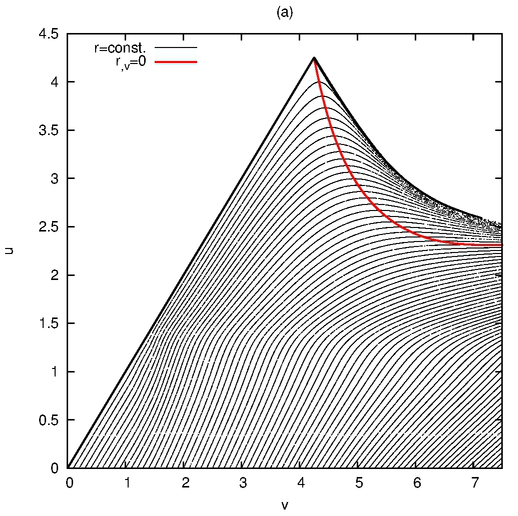}
\includegraphics[scale=0.5]{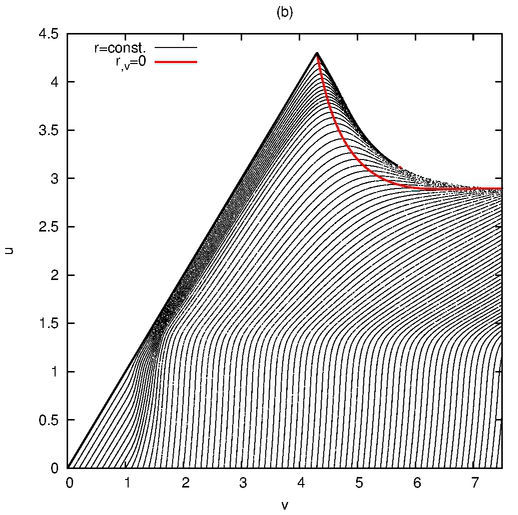}
\caption{(color online).
Lines of constant $r$ in the $(vu)$-plane for $E\overline{MD}$ evolution with a varying dilaton 
field amplitude. The electric and dilatonic coupling constants are equal to $e=0.5$ and $\alpha=-1$, respectively. 
The family 
parameter for the electrically charged scalar field $\tpe_s=0.6$, while for the 
dilaton field (a) $\tpe_k=0.1$ and (b) $\tpe_k=0.3$.}
\label{fig:E-M-DaD-1}
\end{figure}

Let us turn our attention to
the evolutions with a constant amplitude of {\it phantom} electrically charged scalar field equal to $\tpe_s=0.6$. 
For $\alpha=-1$ we considered the values of {\it phantom} dilaton field amplitude $\tpe_k$ equal to $0.1$ and $0.3$. 
The spacetimes emerging from these evolutions are presented in Fig.\ref{fig:E-M-DaD-1}. For all values of $\tpe_k$ the structure of the spacetime resembles Schwarzschild one, i.e., there is an apparent horizon $r_{,v}=0$ coinciding 
with an event horizon at $v\rightarrow\infty$ and surrounding central spacelike singularity. 
The peak, where the line $r=0$ becomes singular practically does not change its $u$-location as $\tpe_k$ increases. On the contrary, the value 
of $u$-coordinate corresponding to the location of the event horizon 
increases while the amplitude of {\it phantom} dilaton field becomes bigger.\\
In the considered case of $\alpha = -1$ the energy momentum tensor (\ref{ten})--(\ref{ten-2}) for $E\overline{MD}$-theory is given by:
\ben \label{taa0}
T_{\mu \nu}(F, \phi) &=& - e^{- 2 \phi} T_{\mu \nu}(F) - T_{\mu \nu}(\phi), \\ \nonumber
T_{\mu \nu}(\phi, F, \psi, \zpsi, A) &=& \tT_{\mu \nu}(\psi, \zpsi, A) + T_{\mu \nu}(F, \phi).
\een
Contrary to the case described by relations (\ref{tt}), where we observe dynamic wormhole spacetimes, now
the {\it phantom} coupling of Maxwell field causes that the outcome is similar to the Schwarzschild
spacetime. It should be emphasized that it happens despite of the influence of {\it phantom} 
dilaton field, which previously prevented black hole formation due to its repulsive nature.
\par
\begin{figure}[p]
\includegraphics[scale=0.5]{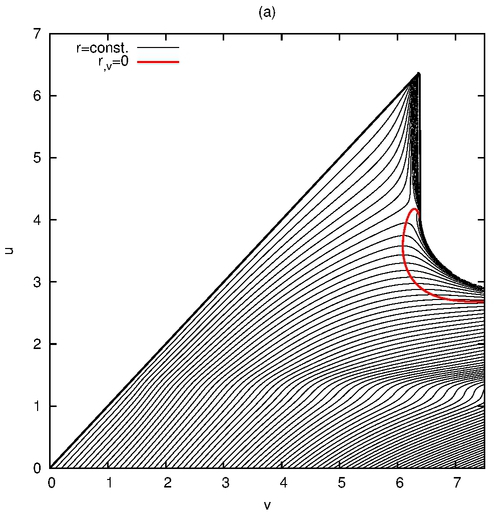}\\
\includegraphics[scale=0.5]{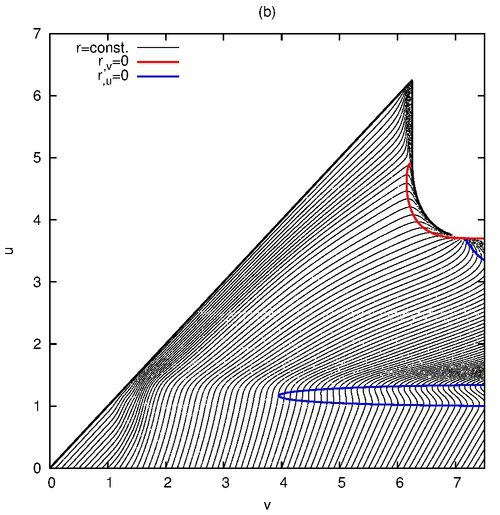}
\caption{(color online).
Lines of constant $r$ in the $(vu)$-plane for the evolution described in Fig.\ref{fig:E-M-DaD-1} with the
dilatonic coupling constant equal to $\alpha=0$ and the family parameter for dilaton field (a) $\tpe_k=0.05$ and (b) $\tpe_k=0.2$.}
\label{fig:E-M-DaD0}
\end{figure}

On the other hand, Fig.\ref{fig:E-M-DaD0} illustrates spacetime structures resulting from 
a collapse in Einstein-{\it phantom Maxwell}-{\it phantom dilaton} 
theory without dilaton coupling, i.e., for $\alpha = 0$. The values of {\it phantom} dilaton
field were set as equal to $0.05$ and $0.2$.
For small values of $\tpe_k$ the spacetime is singular with one apparent horizon $r_{,v}=0$ surrounding singularity 
visible in the domain of integration. Nearby the 
point, where the line $r=0$ becomes singular the apparent horizon almost coincides with the singularity. 
At some point it moves away from the line $r=0$ towards slightly smaller values of 
$v$-coordinate and then settles along constant $u$ for $v\rightarrow\infty$. 
Since it becomes null hypersurface there we may predict 
that its location at large values of advanced time 
indicates the location of the event horizon. 
Although the structure in the vicinity of the point, where the line $r=0$ 
becomes singular (beyond the event horizon) is different from the one presented in Fig.\ref{fig:RNS}a, we 
conclude that it may be interpreted as Schwarzschild-like. The emerging 
object consists of a central spacelike singularity surrounded by one apparent horizon 
coinciding with an event horizon at $v\rightarrow\infty$.
The length of a part, where the apparent horizon is very close 
to $r=0$ decreases with an 
increase of an amplitude of {\it phantom} dilaton field. The location of the point, where 
the line $r=0$ becomes singular slightly decreases in terms of $u$ while the varying amplitude increases.
For larger values of a varying amplitude, naked singularity and an additional open 
loop-shaped branch of an apparent horizon $r_{,u}=0$ appear in the spacetime.\\
Consider now the energy momentum tensor for the studied $\alpha = 0$ case. It implies
\ben \label{taa2}
T_{\mu \nu}(F, \phi) &=& - T_{\mu \nu}(F) - T_{\mu \nu}(\phi), \\ \nonumber
T_{\mu \nu}(\phi, F, \psi, \zpsi, A) &=& e^{2 \phi} \tT_{\mu \nu}(\psi, \zpsi, A) + T_{\mu \nu}(F, \phi).
\een
Having in mind Eq.(\ref{ta0}), the case when no formation of a black hole was indicated, now
we observe that {\it phantom} coupling of Maxwell field provides the singularity in the 
emerging spacetime and the formation of a Schwarzschild-like black hole or naked singulatrity. 
Similarly to the case of $\alpha=-1$ black hole formation takes place in spite of the presence of the 
repulsive {\it phantom} dilaton field.
\par
\begin{figure}[p]
\includegraphics[scale=0.5]{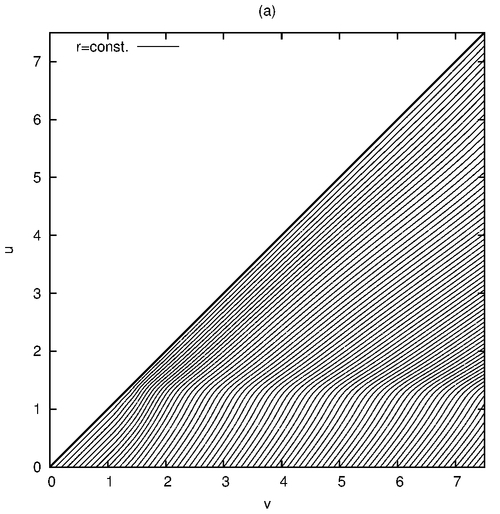}
\includegraphics[scale=0.5]{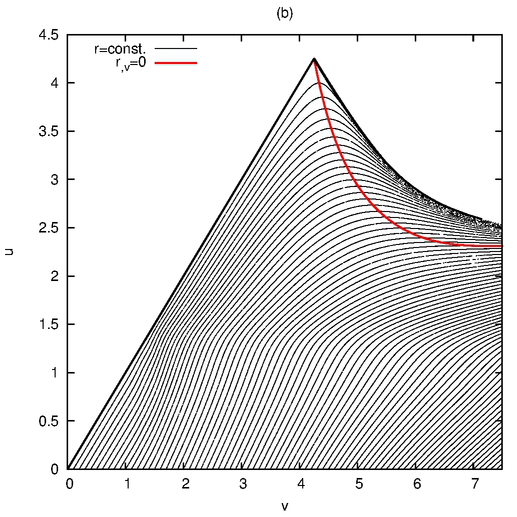}
\caption{(color online).
Lines of constant $r$ in the $(vu)$-plane for $E\overline{MD}$ evolution with a 
varying electrically charged scalar field amplitude. The electric and dilatonic coupling constants are 
equal to $e=0.5$ and $\alpha=-1$, respectively. The family parameter for the dilaton field 
is chosen as $\tpe_k=0.1$, while for the electrically charged scalar field field 
equals to (a) $\tpe_s=0.25$ and (b) $\tpe_s=0.5$, respectively.}
\label{fig:E-M-DaM-1}
\end{figure}

The next set of solutions corresponding to $E\overline{MD}$-theory refers to the computations conducted 
for a constant {\it phantom} dilaton field amplitude $\tpe_k=0.1$ and altering amplitude of {\it phantom} 
counterpart of an electrically charged scalar field. The 
spacetimes ensuing from the collapse for dilatonic coupling constant $\alpha=-1$ and values of an electrically 
charged scalar field $\tpe_s$ equal to $0.25$ and $0.5$ are shown in Fig.\ref{fig:E-M-DaM-1}. For small values 
of an amplitude $\tpe_s$ the spacetime is non-singular, while for larger values it becomes Schwarzschild-type. 
There is one apparent horizon $r_{,v}=0$ indicating the location of an event horizon for $v\rightarrow\infty$. It 
surrounds central spacelike singularity like in Fig.\ref{fig:RNS}a.\\
The damping effect of $e^{-2 \phi}$ in (\ref{taa0}) is dominant at first and the {\it phantom} dilaton field 
plays the most important role.
But with the passage of the increase of the {\it phantom} counterpart of the electrically charged 
scalar field amplitude the situation reverses and this field manages to force black hole formation. 
Just one can conclude that instead of the diminishing factor $e^{-2 \phi}$, the sign of energy momentum 
tensor of $U(1)$-gauge field combined with the increasing
value of the amplitude of the electrically charged scalar field are of the crucial importance for the formation
of Schwarzschild-like black hole.
\par
\begin{figure}[p]
\includegraphics[scale=0.5]{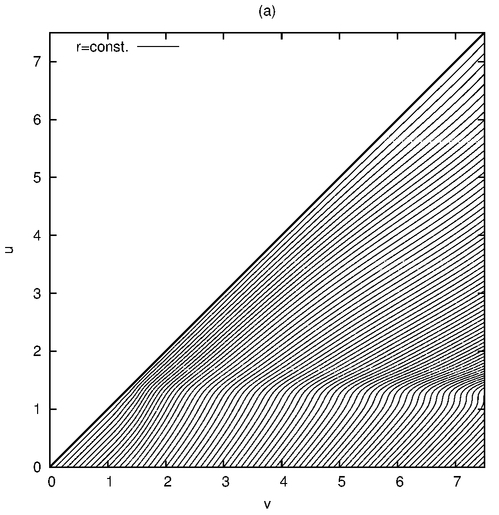}\\
\includegraphics[scale=0.5]{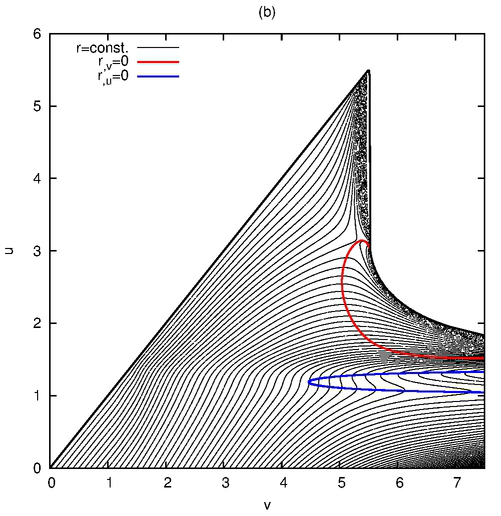}
\caption{(color online).
Lines of constant $r$ in the $(vu)$-plane for the evolution described in Fig.\ref{fig:E-M-DaM-1} with 
the dilatonic coupling constant equal to $\alpha=0$ and the family parameter for the electrically charged 
scalar field field equal to (a) $\tpe_s=0.35$ and (b) $\tpe_s=0.65$.}
\label{fig:E-M-DaM0}
\end{figure}

The diagrams of spacetimes for $\alpha=0$ are illustrated in Fig.\ref{fig:E-M-DaM0}. The amplitude of the 
{\it phantom} electrically 
charged scalar field was equaled to $0.35$ and $0.65$. For small values of $\tpe_s$ the spacetime is non-singular. 
The structures obtained when the amplitude $\tpe_s$ increases 
are Schwarzschild-like with an additional branch of an $r_{,u}=0$ apparent horizon at small values of the retarded time.\\
On the other hand, a close inspection of (\ref{taa2}) reveals
that together with the dominating factor
$e^{2 \phi}$, the opposite sign of the Maxwell term
again occurs decisive for the singularity to appear in the domain under inspection.
\par
As was noticed in the above subsection, the very familiar situation as described in $E M \overline{D}$-theory,
i.e., black hole solution with effectively one event horizon in the case of $\alpha^2 \rightarrow 1$ \cite{gib96},
takes place in $E\overline{MD}$-gravity.
This analytical result obtained in the static case was confirmed in our simulations of the dynamical collapse.

%%%%%%%%%%%%%%%%%%%%%%%%%%%%%%%%%%%%%%%%%%%%%%%%%%%%%%%%%%%%%%%%%%%%%%%%%%%%%%%%%%%%%%%%%%%%%%%%%%%%%%%%%%%%%%%%%%%%%%%%%%%%%%%
\section{Conclusions}

In our paper we studied the problem of the dynamical collapse 
of charged scalar field in the Einstein-Maxwell-dilaton theory with {\it phantom} coupling of the adequate fields.
These fields are allowed to violate the null energy condition. 
The landscape of static and spherically symmetric solutions for the system under consideration abounds in 
eleven different not only asymptotically flat spacetimes related to sixteen various causal structures \cite{cle09}. 
Such a rich collection of solutions aroused our interest especially in the context of the dynamical evolution.
\par
At first, it was revealed that in
Einstein-{\it phantom Maxwell} evolution the structure of the resulting spacetime resembles Schwarzschild-like
spacetime despite of the presence of the electrically charged $U(1)$-gauge field. 
In this case {\it phantom} nature of the electrically charged scalar field forces Schwarzschild-like 
structure formation instead of Reissner-Nordstr\"{o}m-like in the non-{\it phantom} case. What is more, 
the resulting event horizon in the case of $E\overline{M}$ dynamical collapse emerges slightly earlier, in 
terms of the retarded time, than in the case of the dynamical collapse 
in $EM$-theory. Such structure remains in agreement with the one described in non-dynamical case \cite{gib96,cle09}.
\par
Then, taking into account Einstein-{\it phantom dilaton} theory, one observes that for every considered
amplitude of the dilaton field the resulting spacetime is non-singular. There is no formation of
a black hole. Such behaviour is in complete contrast
to the dynamical collapse in Einstein-dilaton (Einstein-scalar) evolution, which leads to the formation of
Schwarzchild-like spacetime \cite{ham96}. 
In this case the {\it phantom} nature of scalar field prevents a singularity 
as well as event horizons formation due to the repulsive character of the field.
\par 
On its turn, in Einstein-{\it phantom Maxwell}-dilaton case there are no qualitative differences among the structures of
the emerging spacetimes for the different values of the coupling constant $\alpha$. One arrives at the
Schwarzschild-like black hole formation, with a single horizon surrounding
spacelike singularity. 
Dilaton field and {\it phantom} counterpart of electrically charged scalar field 
cooperate in Schwarzschild-type structure formation.
These dynamically obtained structures only partially confirm the static solutions 
for $E\overline{M}D$-theory obtained in \cite{cle09}. From a wide variety of possibilities during the dynamical 
evolution the simplest one with only one horizon surrounding spacelike singularity is preferred.
\par
As far as Einstein-Maxwell-{\it phantom dilaton} theory is concerned, the obtained structures 
are far more complicated than in the previous cases. 
First of all, two types of apparent horizons are observed in the spacetime, namely lines of $r_{,u}=0$ and $r_{,v}=0$. 
Moreover, the appearance of an apparent horizon is not necessarily
connected with the emergence of the singularity. For the dilatonic coupling constant equal to $-1$ 
dynamical wormhole spacetimes are observed in case when an amplitude of {\it phantom} dilaton field 
is considerably smaller than an amplitude of an 
electrically charged scalar field. In fact one has a formation of 
the black hole which gains {\it phantom} dilaton field which violates null energy condition.
This phenomenon provides a formation of the wormhole structure.
For $\alpha = 0$ the spacetimes are non-singular. It means that an uncoupled 
{\it phantom} dilaton field prevents black hole formation 
due to its repulsive character.
It seems that in the case of $EM\overline{D}$-collapse the diversity of solutions manifests itself most clearly. Obviously, as was noticed in \cite{cle09,cle11}, multiple horizons separating distinct areas are observed. 
Moreover, as was anticipated in static cases \cite{gib96,cle09}, wormhole structures 
in the case of {\it phantom} scalar field and non-{\it phantom} vector field were obtained 
also during the dynamical collapse. Interestingly, our computations also provided an 
evidence of the existence of the dynamical interconversion between black holes and wormholes, which 
results from the dynamics of the phantom field in the considered spacetime \cite{hay09}.
\par
Finally we note that, in the case of Einstein-{\it phantom Maxwell}-{\it phantom dilaton} theory, 
also both types of the apparent horizons are visible in the domain of integration.
The obtained structures 
correspond to the non-singular or Schwarzschild-type spacetimes or they contain naked singularities. 
Naked singularities mainly appear when the amplitudes of both collapsing fields are equal.
On the other hand, the dynamical collapse results in Schwarzschild-like spacetime when an amplitude 
of the dilaton field is considerably smaller than the other one. Hence it may be stated that 
the considered $E\overline{MD}$-evolution 
again leads to the simplest structure with a singularity surrounded by an event horizon among those 
presented in \cite{cle09}.
\par
During our numerical simulations of a dynamical collapse in {\it phantom}-$EMD$ system we obtained an 
extensive collection of the spacetime structures. They are summarized in Table \ref{tab:summary} 
together with the structures emerging from the collapse without {\it phantom} coupling. An analysis 
of energy-momentum tensors and the nature of forces acting in the considered system 
combined with the study of the emerging spacetime structures, led us to the general conclusions concerning 
the role of particular fields and the dilatonic coupling during the collapse in question.
\par
It turned out that the non-{\it phantom} dilaton field supports the formation of 
Schwarzschild-type singular spacetimes. On the other hand, its {\it phantom} 
counterpart suppresses black hole formation due to its strongly repulsive nature. 
It also facilitates the formation of multiple horizons not necessarily connected with the singularity.
The non-{\it phantom} Maxwell field seems to be less repulsive than {\it phantom} dilaton field 
and thus it favors Reissner-Nordstr\"{o}m-type spacetime as an outcome of the
dynamical collapse. What is more, {\it phantom} Maxwell field, similarly to the dilaton field, supports the 
creation of the simplest singular, Schwarzschild-like structure despite its {\it phantom} coupling to gravity.\\ 
Considering the role of a complex scalar field during the collapse, it seems to act just as a carrier of the 
Maxwell field. It does not influence the obtained structures distinctly. Such situation is in agreement with 
the previously examined simpler case of the collapse of electrically charged scalar field in Einstein 
gravity \cite{ore03}.
\par
Considering the dilatonic coupling it may be revealed that $\alpha=-1$ diminishes the influence of Maxwell field 
on the evolution and thus enhances the role of the dilaton field in its course. The dilatonic coupling 
constant equal to zero acts the other way round, i.e., it reduces the role of dilaton field during the 
collapse and indirectly enlarges the meaning of Maxwell field by acting on the complex electrically charged scalar field.

\begin{table*}[t]
\caption{Spacetime structures emerging from the {\it phantom} gravitational collapse. The overlining 
indicates {\it phantom} field. The structures of spacetimes resulting from the evolutions in non-{\it phantom} gravity are summarized at the bottom for comparison. Symbols: $\mathcal{A}$ -- attractive, $\mathcal{R}$ -- repulsive, $ns$ -- non-singular spacetime, $S$ -- Schwarzschild-type spacetime, $RN$ -- Reissner-Nordstr\"{o}m-type spacetime, $WH$ -- wormhole, $NS$ -- naked singularity. Superscripts: $^\ast$ -- additional horizons in spacetime, $^{\left(2\right)}$ -- two-stage collapse. Arrows: $\uparrow$ -- an increase of a particular amplitude, $\rightarrow$ -- the direction of changes while the amplitude varies.}
\begin{ruledtabular}
\begin{tabular}{c|c|c|c|c|c|c|c}
\multirow{2}{*}{Evolution} & \multicolumn{2}{c|}{Forces} & Dilatonic & Energy-momentum & \multirow{2}{*}{$\tpe_s = \tpe_k\uparrow$} & $\tpe_s=const.$ & $\tpe_k=const.$\\ \cline{2-3}
& $M/\overline{M}$ & $D/\overline{D}$ & coupling constant & tensor & & $\tpe_k\uparrow$ & $\tpe_s\uparrow$\\
\hline
& & & & & & &\\

$E\overline{M}$ & $\mathcal{A}$ & $-$ & $-$ & $\tT_{\mu\nu}\left(\psi\right)-T_{\mu\nu}\left(F\right)$ & $-$ & $-$ & $ns\rightarrow S$\\
& & & & & & &\\

$E\overline{D}$ & $-$ & $\mathcal{R}$ & $-$ & $-T_{\mu\nu}\left(\phi\right)$ & $-$ & $ns$ & $-$\\
& & & & & & &\\

\multirow{2}{*}{$E\overline{M}D$} & \multirow{2}{*}{$\mathcal{A}$} & \multirow{2}{*}{$\mathcal{A}$} & $\alpha=-1$ & $\tT_{\mu\nu}\left(\psi\right)-e^{-2\phi}T_{\mu\nu}\left(F\right)+T_{\mu\nu}\left(\phi\right)$ & \multirow{2}{*}{$ns\rightarrow S$} & \multirow{2}{*}{$S\rightarrow S^{\left(2\right)}$} & \multirow{2}{*}{$S\rightarrow S^{\left(2\right)}$}\\
& & & $\alpha=0$ & $e^{2\phi}\tT_{\mu\nu}\left(\psi\right)-T_{\mu\nu}\left(F\right)+T_{\mu\nu}\left(\phi\right)$ & & &\\
& & & & & & &\\

\multirow{2}{*}{$EM\overline{D}$} & \multirow{2}{*}{$\mathcal{R}$} & \multirow{2}{*}{$\mathcal{R}$} & $\alpha=-1$ & $\tT_{\mu\nu}\left(\psi\right)+e^{-2\phi}T_{\mu\nu}\left(F\right)-T_{\mu\nu}\left(\phi\right)$ & $ns\rightarrow ns^\ast$ & $WH\rightarrow NS\rightarrow ns$ & $ns\rightarrow WH$\\
& & & $\alpha=0$ & $e^{2\phi}\tT_{\mu\nu}\left(\psi\right)+T_{\mu\nu}\left(F\right)-T_{\mu\nu}\left(\phi\right)$ & $ns\rightarrow ns^\ast$ & $ns^\ast$ & $ns\rightarrow ns^\ast$\\
& & & & & & &\\

\multirow{2}{*}{$E\overline{MD}$} & \multirow{2}{*}{$\mathcal{A}$} & \multirow{2}{*}{$\mathcal{R}$} & $\alpha=-1$ & $\tT_{\mu\nu}\left(\psi\right)-e^{-2\phi}T_{\mu\nu}\left(F\right)-T_{\mu\nu}\left(\phi\right)$ & $ns\rightarrow NS$ & $S$ & $ns\rightarrow S$\\
& & & $\alpha=0$ & $e^{2\phi}\tT_{\mu\nu}\left(\psi\right)-T_{\mu\nu}\left(F\right)-T_{\mu\nu}\left(\phi\right)$ & $ns\rightarrow NS^\ast$ & $S\rightarrow NS^\ast$ & $ns\rightarrow S^\ast$\\
& & & & & & &\\

\hline
& & & & & & &\\

$EM$ \cite{ore03} & $\mathcal{R}$ & $-$ & $-$ & $\tT_{\mu\nu}\left(\psi\right)+T_{\mu\nu}\left(F\right)$ & $-$ & $-$ & $ns\rightarrow RN$\\
& & & & & & &\\

$ED$ \cite{ham96} & $-$ & $\mathcal{A}$ & $-$ & $T_{\mu\nu}\left(\phi\right)$ & $-$ & $ns\rightarrow S$ & $-$\\
& & & & & & &\\

\multirow{2}{*}{$EMD$ \cite{bor11}} & \multirow{2}{*}{$\mathcal{R}$} & \multirow{2}{*}{$\mathcal{A}$} & $\alpha=-1$ & $\tT_{\mu\nu}\left(\psi\right)+e^{-2\phi}T_{\mu\nu}\left(F\right)+T_{\mu\nu}\left(\phi\right)$ & $ns\rightarrow S$ & $S\rightarrow S^{\left(2\right)}$ & $S\rightarrow S^{\left(2\right)}$\\
& & & $\alpha=0$ & $e^{2\phi}\tT_{\mu\nu}\left(\psi\right)+T_{\mu\nu}\left(F\right)+T_{\mu\nu}\left(\phi\right)$ & $ns\rightarrow RN$ & $RN\rightarrow RN^{\left(2\right)}$ & $RN\rightarrow RN^{\left(2\right)}$\\
& & & & & & &\\

\end{tabular}
\end{ruledtabular}
\label{tab:summary}
\end{table*}

\begin{acknowledgments}
AN was supported by Human Capital Programme of European Social Fund sponsored by European Union.\\                   
MR was partially supported by the grant of the National Science Center $2011/01/B/ST2/00408$.
\end{acknowledgments}
%%%%%%%%%%%%%%%%%%%%%%%%%%%%%%%%%%%%%%%%%%%%%%%%%%%%%%%%%%%%%%%%%%%%%%%%%%%%%%%%%%%%%%%%%%%%%%%%%%%%%%%%%%%%%%%%%%%%%%%%%%%%%

%%%%%%%%%%%%%%%%%%%%%%%%%%%%%%%%%%%%%%%%%%%%%%%%%%%%%%%%%%%%%%%%%%%%%%%
%\begin{appendix}

%\section{Irred   } 
%\label{irtf}
%\end{appendix}
%%%%%%%%%%%%%%%%%%%%%%%%%%%%%%%%%%%%%%%%%%%%%%%%%%%%%%%%%%%%%%%%%%%%%%%%


%%%%%%%%%%%%%%%%%%%%%%%%%%%%%%%%%%%%%%%%%%%%%%%%%%%%%%%%%%%%%%%%%%%%%%%%%%%%%%%%
%%%%%%%%%%%%%%%%%%%%%%%%%%%%%%%%%%%%%%%%%%%%%%%%%%%%%%%%%%%%%%%%%%%%%%%%%%%%%%%

%%%%%%%%%%%%%%%%%%%%%%%%%%%%%%%%%%%%%%%%%%%%%%%%%%%%%%%%%%%%%%%%%%%%%%%%%%%%%%%%%%
%%%%%%%%%%%%%%%%%%%%%%%%%%%%%%%%%%%%%%%%%%%%%%%%%%%%%%%%%%%%%%%%%%%%%%%%%%%%%%%%%%%
\end{document}